\documentclass[preprint2]{aastex}

\usepackage{amsmath} 
\usepackage{natbib}

\shorttitle{BEST\,II Variable Star Catalog I}
\shortauthors{Fruth et al.}

\begin{document}

\title{The Berlin Exoplanet Search Telescope II. \\
Catalog of Variable Stars. \\
I. Characterization of Three Southern Target Fields.}

\author{T.~Fruth\altaffilmark{1}, J.~Cabrera\altaffilmark{1}, R.~Chini\altaffilmark{3,4}, Sz.~Csizmadia\altaffilmark{1}, C.~Dreyer\altaffilmark{1,6}, P.~Eigm\"uller\altaffilmark{1}, A.~Erikson\altaffilmark{1}, P.~Kabath\altaffilmark{1,2}, S.~Kirste\altaffilmark{1}, R.~Lemke\altaffilmark{3}, M.~Murphy\altaffilmark{5}, T.~Pasternacki\altaffilmark{1}, H.~Rauer\altaffilmark{1,6}, and R.~Titz-Weider\altaffilmark{1}} 

\altaffiltext{1}{Institut f\"ur Planetenforschung, Deutsches Zentrum f\"ur Luft- und Raumfahrt, 
Rutherfordstr.~2, 12489~Berlin, Germany}
\altaffiltext{2}{European Southern Observatory, Alonso de C\'ordova~3107, Vitacura, Casilla~19001, Santiago~19, Chile}
\altaffiltext{3}{Astronomisches Institut, Ruhr-Universit\"at Bochum, 44780~Bochum, Germany}
\altaffiltext{4}{Instituto de Astronom\'{\i}a, Universidad Cat\'{o}lica del Norte, Antofagasta, Chile}
\altaffiltext{5}{Depto.~F\'isica, Universidad Cat\'olica del Norte, PO~1280, Antofagasta, Chile}
\altaffiltext{6}{Zentrum f\"ur Astronomie und Astrophysik, Technische Universit\"at Berlin, 10623~Berlin, Germany}
\email{thomas.fruth@dlr.de}

\begin{abstract}
A photometric survey of three Southern target fields with BEST\,II yielded the detection of~2{,}406 previously unknown variable stars and an additional 617 stars with suspected variability. This study presents a catalog including their coordinates, magnitudes, light curves, ephemerides, amplitudes, and type of variability. In addition, the variability of 17 known objects is confirmed, thus validating the results. The catalog contains a number of known and new variables that are of interest for further astrophysical investigations, in order to, e.g., search for additional bodies in eclipsing binary systems, or to test stellar interior models.

Altogether, 209{,}070 stars were monitored with BEST\,II during a total of 128 nights in 2009/2010. The overall variability fraction of 1.2--1.5\% in these target fields is well comparable to similar ground-based photometric surveys. Within the main magnitude range of $R\in\left[11,17\right]$, we identify 0.67(3)\% of all stars to be eclipsing binaries, which indicates a completeness of about one third for this particular type in comparison to space surveys.  
\end{abstract}

\keywords{techniques: photometric --- binaries: eclipsing ---  stars: variables}

\section{Introduction}
The detailed study of variable stars is essential to astronomy, since it allows for the determination of stellar parameters such as mass, radius, luminosity, or temperature, as well as to study internal and external processes of stars, their composition, structure, and evolution. New detections not only broaden the statistical sample of variable stars, but are also important to gain further knowledge about the different processes that cause stellar variability. 

The Berlin Exoplanet Search Telescopes, BEST \citep{Rauer2004} and BEST\,II \citep{Kabath2009}, are small-aperture, wide-field telescopes that are primarily used as a ground-based support for the CoRoT space mission \citep{Baglin2006}. Their observations help to exclude false positives from the list of transiting planetary candidates that are identified in time series from the satellite \citep{Deeg2009,Rauer2010,Csizmadia2011}. In addition, long-term photometric monitoring enables a precise characterization of stellar variability in CoRoT target fields \citep{Karoff2007,Kabath2007,Kabath2008,Kabath2009,Kabath2009a,Fruth2012,Klagyivik2013} and beyond \citep{Pasternacki2011}.

During time periods not required for the regular CoRoT support, BEST\,II started an independent transit survey in 2009. Up to now, seven southern target fields have been monitored. While the analysis and follow-up of planetary candidates is ongoing, this paper presents a photometric analysis of stellar variability within the first three target fields of this survey.

The paper is organized as follows: Section~\ref{sec:telescope} describes the telescope configuration and the observational data. Section~\ref{sec:reduction} summarizes the data reduction and photometric analysis, while the search for stellar variability is outlined in Section~\ref{sec:variability}. The results are presented in a large catalog of variable stars in Section~\ref{sec:results}, which includes a photometric classification of the variability type, ephemerides, a comparison with literature results for known cases, and a discussion on its limitations. Finally, Section~\ref{sec:summary} briefly summarizes the paper.

\section{Telescope and Observations}\label{sec:telescope}
BEST\,II is located at the Observatorio Cerro Armazones, Chile. Since 2007, it is operated continuously by the Institute of Planetary Research of the German Aerospace Center (DLR) in robotic mode from Berlin. 

\begin{table*}[t]\centering\small
\caption{BEST\,II Target Field Information\label{tab:fields}}
 \begin{tabular}{lc@{\ }rrcrrr}
\tableline\tableline
Field & \multicolumn{2}{c}{Coordinates (J2000.0)} & \multicolumn{1}{c}{Season} & \multicolumn{1}{c}{Nights} & \multicolumn{1}{c}{Fra-} & \multicolumn{2}{c}{Stars} \\
 & $\alpha$ & \multicolumn{1}{c}{$\delta$} &&& \multicolumn{1}{r}{mes} & Total & $\sigma\leq 0.01^m$ 
\\\tableline
F17 & $14^h 24^m 29^s$ & $-54\degr 07\arcmin 20\arcsec$ & 20/04/09--22/07/09 & 39 & 2{,}259 &  68{,}317 & 3{,}170  \\
F18 & $22^h 52^m 00^s$ & $-44\degr 12\arcmin 00\arcsec$ & 19/08/09--27/10/09 & 27 & 2{,}266 &  13{,}551 & 427  \\
F19 & $16^h 26^m 00^s$ & $-56\degr 12\arcmin 00\arcsec$ & 24/03/10--21/09/10 & 62 & 2{,}855 & 127{,}202 & 10{,}120 \\\tableline
\end{tabular}
\tablecomments{Shown are the center coordinates, the time range between the first and last observing night, the number of good photometric nights with observations within this range, the number of acquired frames, and the number of total/low-noise light curves for each target field.}
\end{table*}

\begin{figure*}[t]\centering
  \includegraphics[width=.3\linewidth]{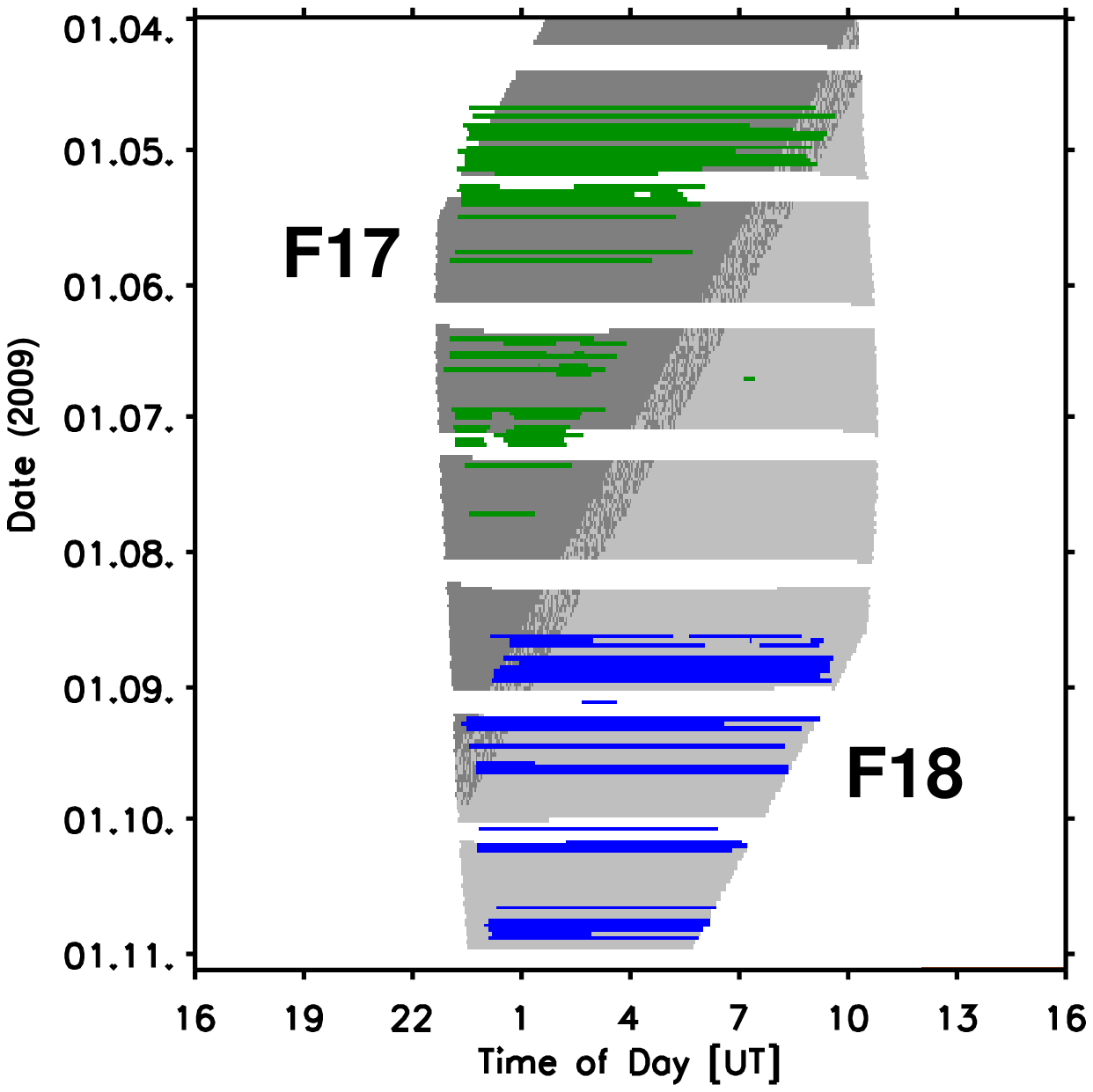}\hspace{2cm} 
  \includegraphics[width=.3\linewidth]{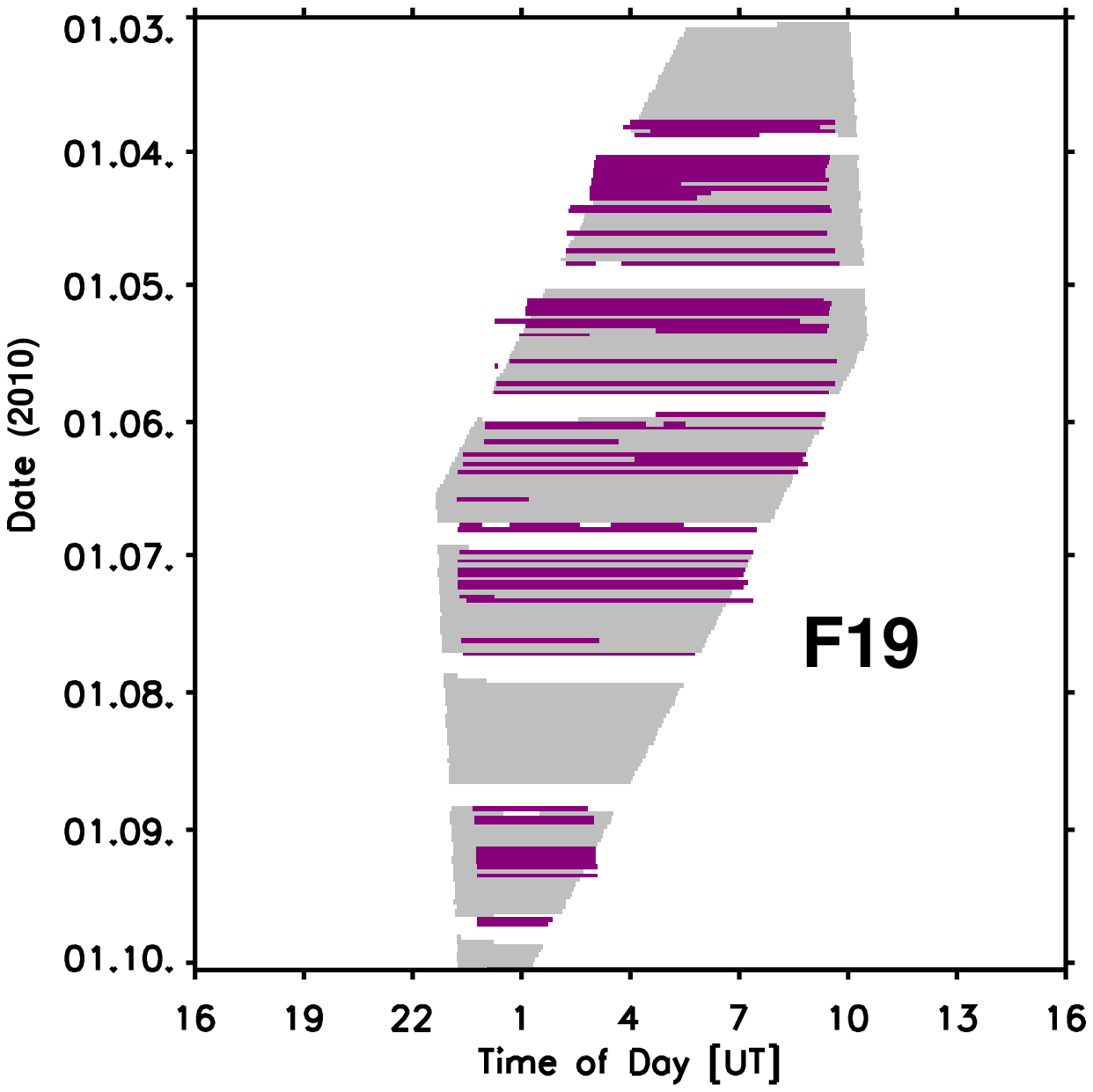}\vspace{5mm}
\figcaption{BEST\,II field observations during Chilean winters 2009 and 2010. Times of observations are shown for the fields F17 (green), F18 (blue), and F19 (violet). For comparison, gray areas indicate times with an \textit{optimal} astronomical visibility, i.e., corresponding to each respective target field being observable at least~$30\degr$ above the horizon, the Sun below~$-8\degr$, and a Moon phase of less than~90\% \citep[cf.][]{Rauer2008}. \label{fig:best2obs}}
\end{figure*}

The system consists of a 25\,cm Baker-Ritchey-Chr\'etien telescope with a focal ratio of f/5.0 and a $1\fdg7$\,$\times$\,$1\fdg7$ wide field of view (FOV). The photometric data presented here were obtained with a~\mbox{4k\,$\times$\,4k} Finger Lakes Instrumentation CCD (IMG-16801E1) in white light, i.e., without any photometric filter. The CCD is most sensitive at $\lambda\approx 650$~nm, and the photometric system is roughly comparable to the Johnson $R$-band. The pixel size of 9\,$\mu$m corresponds to an angular resolution of $1\farcs5$\,pixel$^{-1}$.

Three target fields, named F17, F18, and F19, have been monitored intensively with BEST\,II in 2009/2010. They have been selected by maximizing the total observing time weighted against the average airmass \citep[][Equation~5]{Rauer2008} for each period of planned observations. For selecting F19, the simulation was complemented by additionally maximizing the number of target stars suitable for transit search (main sequence, less than 1\% contamination). The stellar density of the fields differs significantly, since F17 and F19 are located close to the galactic plane ($\left|b\right|\approx 5\degr$), while F18 is well outside ($b=-61\degr$). The respective center coordinates of all three target fields are given in Table~\ref{tab:fields}, which also lists the number of frames and light curves obtained in each pointing direction. 

In total, BEST\,II collected 7{,}380 scientific frames and recorded light curves for 209{,}070 stars in these target fields. For the first field, F17, BEST\,II observations cover 39 photometric nights between 20/04/2009 and 22/07/2009. Field F18 was observed for 27 nights between 19/08/2009 and 27/10/2009, and field F19 for 62 nights between 24/03/2010 and 21/09/2010 (see Figure~\ref{fig:best2obs}). When the observing coverage (colored areas in Figure~\ref{fig:best2obs}) is related to the maximum available night time during these periods (gray areas), this corresponds to an average duty cycle of 35\% for F17 and F19, and 38\% for F18. Target fields F17 and F18 were observed with an exposure time of 120\,s, while 300\,s exposures were taken for F19; the typical cadence between two adjacent measurements is 2.5~minutes.

\section{Data Reduction}\label{sec:reduction}
The methods used here to obtain photometric time series from raw scientific images are part of a dedicated automated pipeline that has been applied before to various BEST/BEST\,II data sets \citep{Karoff2007,Kabath2007,Kabath2008,Kabath2009,Kabath2009a,Rauer2010,Pasternacki2011,Fruth2012,Klagyivik2013}.

Calibration frames (bias, dark, flat) were recorded together with the observations and used in a standard reduction of instrumental effects. In order to increase the photometric precision in crowded fields, we apply the image subtraction algorithm \citep{Alard1998,Alard2000}. For that, the calibrated scientific images are aligned to a common coordinate system, and the best~\mbox{$\sim$\,20--40} scientific frames are stacked to a reference image. The latter is then fitted to individual frames and subtracted.

Simple unit-weight aperture photometry is used to extract both the flux from the reference frame and the relative flux in each subtracted frame. A standard radius of 5~pixels was chosen for target fields F17 and F19, while 7~pixels were used in the reduction of the less dense F18 field. An adjacent annulus up to an outer radius of 20~pixels is used for an estimation of the background flux. In order to remove global flux variations in the data, e.g., due to weather or nightly variations of the sky transperancy, a comparison star is calculated out of some thousand light curves with the smallest photometric noise in each data set and then subtracted from each light curve.

Finally, all stars are matched with the UCAC3 catalog \citep{Zacharias2010} in order to assign equatorial coordinates and to adjust instrumental magnitudes to a standard magnitude system. The astrometric calibration is obtained using the routines \verb+grmatch+ and \verb+grtrans+ by \citet{Pal2006}; for the three data sets presented here, it achieves a match for $\sim$\,85\% of the stars with an average residual of $0\farcs2$--$0\farcs3$. The magnitude calibration is obtained by shifting each data set by the median difference between all instrumental magnitudes and their respective catalog value (R2MAG of UCAC3). Since the photometric systems are comparable but not identical, this calibration yields an accuracy of $\sim$\,0.3--0.5~mag.

\begin{figure}[htpc]\centering 
  \includegraphics[width=\linewidth]{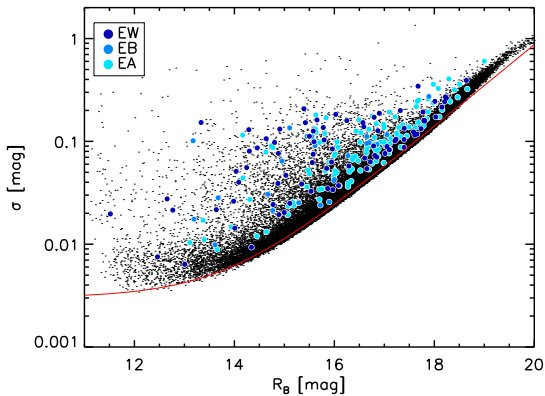}
\figcaption{Photometric standard deviation~$\sigma$ in unbinned BEST\,II data as a function of the instrumental magnitude $R_B$ (example for target field F17). Blue dots indicate binaries in the field (cf.~Section~\ref{sec:results}), while the red line shows the minimum noise level $\sigma_\textrm{min}(R_B)$ in this data set. \label{fig:rmsplot}}
\end{figure}

Brightness variation, however, can be measured to a much higher precision with BEST\,II. For the brightest stars in each data set, it obtains a noise level of $\sim$\,3\,mmag in unbinned data over the whole observing season. As an example, Figure~\ref{fig:rmsplot} shows the standard deviation~$\sigma$ as a function of stellar magnitude for target field F17. Overall, BEST\,II obtained measurements with mmag-precision ($\sigma\leq 0.01$~mag) for 13{,}717 stars in the three target fields presented here (cf.~Table~\ref{tab:fields}).

\section{Search for Variability}\label{sec:variability}
The three data sets were searched for stellar variability using a combination of the $J$-index \citep{Stetson1996} and an analysis of variance period search \citep[AoV;][]{Schwarzenberg-Czerny1996}. The method and its parameters have been described in detail by \citet{Fruth2012}.

First, the $J$-index is used to identify light curves that are clearly \textit{not} variable. Using a limit $J<J_\textrm{max}$, we excluded 93{,}583 non-variable stars; thereby, $J_\textrm{max}$ was set to 0.1 for the crowded target fields F17 and F19, while 0.05 was chosen for the less dense F18 field. Second, the remaining 115{,}487 light curves were each fitted with seven harmonics having fundamental periods within the range of 0.05--100\,days (F17, F19) and 0.05--80\,days (F18), respectively. Third, a modified ranking parameter $q$ was calculated from the AoV statistic, whereby lower weights are given to periodic variability that is encountered in many light curves. 

Finally, a total of 5{,}480 light curves with $q\geq 10$ were inspected visually for stellar variability (for the respective star count within each target field, see Table~\ref{tab:varstars}). This mainly included reviewing the overall signal-to-noise ratio (SNR), checking individual nights or events for systematic effects, inspecting the stellar neighborhood for possible sources of contamination, and analyzing alternative period solutions. The latter step lead us to adjust the initial ephemerides in several cases -- usually to multiples of the AoV period, e.g., due to primary and secondary eclipses of (slightly) different depths.

For some cases, the angular resolution of BEST\,II is not sufficient to fully separate the light of adjacent stars, and the photometric apertures overlap. Thus, variability of the same shape and period can be detected in multiple light curves. In order to constrain the source of variability, we carefully checked each of these \textit{contaminated} objects using a smaller photometric aperture (3\,Px radius). 

Stellar crowding also leads to an overestimation of the instrumental magnitude and an underestimation of the amplitude of variability. In order to assess this effect quantitatively, we calculated the flux fraction within each photometric aperture $i$ that does \textit{not} originate from the respective target~$i$ as
\begin{equation}
  \gamma_i = 1-\frac{g_{ii} f_i}{\sum_j{g_{ij} f_j}} = 1-\frac{g_{ii}}{\sum_j{g_{ij} 10^{0.4\left(m_i-m_j\right)}}},
\end{equation}
whereby $f_j$ denotes the flux of star $j$, $m_j$ its magnitude, and~$g_{ij}$ describes the geometric integral of its PSF within the target aperture~$i$ (normalized, i.e., $g_{ij}\in\left[0,1\right]$). The calculation of $\left(g_{ij}\right)$ assumes Gaussian PSFs (3\,Px FWHM) and circular apertures (radii as used for photometry), whereas $R$ magnitudes from the NOMAD catalog \citep{Zacharias2004} are used to estimate $\left(m_j\right)$. Due to missing catalog entries or magnitudes, the calculation of~$\gamma$ can be inaccurate or even fail in some cases. Also, its accuracy is affected by systematic effects such as the difference in the two photometric systems (BEST\,II and catalog), deviations of the PSF from a Gaussian shape, or long-term stellar variability. Thus, the automatically calculated~$\gamma$ value in Table~\ref{tab:varcat} should only be taken as a first-order approximation.

\section{Results}\label{sec:results}
The visual inspection revealed 2{,}406 stars with clear and previously unknown variability. In addition to that, we confirm the known variability of~17 stars and suspect further~617 objects to be variable. For the latter group, the variability itself, its type, and/or period could not be determined without ambiguity. Predominantly, these are objects that show brightness variations close to the noise level of their light curve. The numbers of detected, known, and suspected variables within each target field are summarized in Table~\ref{tab:varstars}. 

\begin{table*}[tp]\centering\small
\caption{Number of Stars Identified as Variables in BEST\,II target fields F17--F19. \label{tab:varstars}}
\begin{tabular}{lrrrrrrrr}
\tableline\tableline
\textsc{Field} & \multicolumn{3}{c}{\dotfill\textsc{Stars}\dotfill} && \multicolumn{3}{@{}c}{\dotfill\textsc{Variable Stars}\dotfill} \\
         & \multicolumn{1}{c}{\textsc{Total}} & \multicolumn{1}{c}{$J\geq J_\textrm{max}$} & \multicolumn{1}{c@{}}{$q\geq 10$} && \multicolumn{1}{@{}r@{}}{\textsc{Known}} & \multicolumn{1}{c}{\textsc{New}} & \multicolumn{1}{@{}l}{\textsc{Suspected}} \\
\tableline
F17                    &  68{,}317 &  20{,}965 $^{(31\%)}$ & 1{,}126\,$^{(1.7\%)}$ &&  2&     639 $^{(0.94\%)}$ & 179\,$^{(0.26\%)}$ \\
F18                    &  13{,}551 &   5{,}399 $^{(40\%)}$ &     176\,$^{(1.3\%)}$ &&  4&      13 $^{(0.10\%)}$ &   7\,$^{(0.05\%)}$ \\
F19                    & 127{,}202 &  89{,}123 $^{(70\%)}$ & 4{,}178\,$^{(3.3\%)}$ && 11& 1{,}754 $^{(1.38\%)}$ & 431\,$^{(0.34\%)}$ \\\tableline
\textsc{Total}         & 209{,}070 & 115{,}487 $^{(55\%)}$ & 5{,}480\,$^{(2.6\%)}$ && 17& 2{,}406 $^{(1.15\%)}$ & 617\,$^{(0.30\%)}$ \\
\tableline
\end{tabular}
\tablecomments{For each field, the table gives the total number of light curves, the number of light curves selected for variable star search (with $J\geq J_\textrm{max}$), the number of light curves that are finally analyzed visually (with variability parameter $q\geq 10$, cf.~\citealt{Fruth2012}), and the number of known, new, and suspected variable stars. Numbers in brackets give the relative fraction compared to the total count.}
\end{table*}
\begin{table*}[tp]\centering\scriptsize
\caption{Star Counts per Variability Class. \label{tab:varstartypes}}
\begin{tabular}{l c@{ }c@{ }c@{ }c@{ }c@{ }c@{ }c@{ }c@{ }c@{ }c@{ }c@{ }c@{ }c} 
\tableline\tableline
      & EA       & EB       & EW       & EW/DSCT  & DSCT     & RR      & CEP                      & GDOR  & ROT                        & SR       & LP       & VAR       \\\tableline
 F17  &  83  (9) &  24  (7) & 119  (6) &  41 (11) &  46 (23) & 40 (12) & 29                  (9)  & 1 (0) &  77                   (38) &  14  (3) & 156 (56) &  11   (5) \\
 F18  &   0  (1) &   0  (1) &   3  (0) &   0  (0) &   0  (0) &  6  (0) &  0                  (0)  & 0 (0) &   4                    (4) &   0  (0) &   1  (0) &   3   (1) \\
 F19  & 281 (30) & 127 (28) & 317 (22) &  97 (49) & 118 (31) & 53  (9) & 59\tablenotemark{*} (6)  & 0 (0) & 119\tablenotemark{**} (39) & 161 (46) & 187 (22) & 246 (149) \\\tableline
Total & 364 (40) & 151 (36) & 439 (28) & 138 (60) & 164 (54) & 99 (21) & 87\tablenotemark{*} (15) & 1 (0) & 200\tablenotemark{**} (81) & 175 (49) & 344 (78) & 260 (155) \\
\tableline
\end{tabular}
\tablenotetext{*}{\scriptsize Including one CEP/EA}
\tablenotetext{**}{\scriptsize Including one ROT/EA and one ROT/EB}
\tablecomments{Given are the numbers of known and newly detected variable stars for each field and variability class (suspected variables in brackets). Crowded cases are only counted once.}
\end{table*}

\subsection{Classification}\label{sec:classification}
Detected variable stars are assigned variability types following \citet{Sterken1996} and the classification scheme of the General Catalog of Variable Stars \citep[GCVS;][]{Samus2009}. The identification is solely based on photometry, i.e., it depends on the shape, amplitude, and period of the brightness variation.

The following classes could be identified:
\begin{itemize}
	\item \textbf{Eclipsing binary systems.} Light curves with clear eclipses and almost no variation in between are classified as Algol-type binaries (EA; prototype~$\beta$~Per). For systems with ellipsoidal components, phase variations are significant and hinder an exact determination of the beginning/end of eclipses (EB type;~$\beta$~Lyr). At orbital periods below one day, both objects are in contact, eclipses are of equal depth and are fully blended with the phase variation (EW type; W UMa).
	\item \textbf{Pulsating variable stars.} 
	From photometry, the following pulsating types could be identified:  $\delta$~Scuti variables (DSCT; periods $p\leq 0.2$\,days), RR Lyrae (RR; $p=0.2$--1\,day, characteristic shape), Cepheids (CEP; $p>1$\,day, amplitudes 0.01--2\,mag), Gamma Doradus stars (GDOR), and semi-regular variables (SR; $p\geq 20$\,days with irregularities).
	\item \textbf{Rotating variable stars (ROT).} 	
	Stellar rotation can introduce flux variations, e.g., due to stellar spots, magnetic fields, or ellipsoidal components. Since BEST\,II photometry alone is usually not sufficient to distinguish these cases, they are grouped under a general "ROT" classification.
	\item \textbf{Long periodic variables (LP).} Non-periodic variables or stars variable on time scales comparable to/larger than the observational coverage are named LP.	
	\item \textbf{Inconclusive Cases (VAR).} Stars showing clear variability that cannot be assigned a type according to the classification scheme from photometry; further observations are needed to better constrain the physical origin of variability.
\end{itemize}

Details on how many stars have been found in each variability class and target field are given in Table~\ref{tab:varstartypes}. In total, 
954 (plus~104 suspected) eclipsing binaries 
could be identified, 
527~(139) pulsators, 
200~(81) stars with rotational modulation, 
344~(78) long periodic variables, 
and 398~(215) with other types of variability. 

\begin{table*}[htp]\scriptsize\centering
\caption{Known variable stars in data sets F17--F19. \label{tab:varstars:known}}
\begin{tabular}{l@{\ \ }l@{\ \ }cr@{}l@{}r@{}l@{\,}c@{}cl}
\tableline\tableline
 \multicolumn{2}{c}{\dotfill\textsc{Identifier}\dotfill} & $R_B$ & \multicolumn{4}{c}{\dotfill\textsc{Period} $P$ [d]\dotfill} & \multicolumn{2}{c}{\dotfill\textsc{Classification}\dotfill} &  \textsc{Reference} \\
BEST\,II & Ref. & [mag] & \multicolumn{2}{c}{BEST\,II} & \multicolumn{2}{c}{Ref.} & BEST\,II & Ref.  \\\hline\\[-.8em]
F17\_03458  & ASAS J142013-5339.9 & 11.6 & 3&.329(4)  &   3&.31   & ROT & ROT  & \citet{Kiraga2012} \\
F17\_32682  & ASAS J142428-5416.0 & 12.5 &  &--       & 323&.7    & LP  & MISC & \citet{Pojmanski2002} \\
F18\_02074  & ASAS J224935-4341.2 & 11.9 &  &--       &  68&.25   & LP  & MISC & \citet{Pojmanski2002} \\
F18\_03793  & YZ Gru              & 16.5 & 0&.6976(7) &   0&.6974 & RR  & RRAB & \citet{Meinunger1979} \\
F18\_05548  & BE Gru              & 14.1 & 0&.6055(2) &   0&.6054 & RR  & RRAB & \citet{Meinunger1979} \\
F18\_08895  & AD Gru              & 15.5 & 0&.7592(5) &   0&.7592 & RR  & RRAB & \citet{Meinunger1979} \\
F19\_000499 & FV Nor              & 12.6 &  &--       &    &--    & LP  & LP?  & \citet{Hoffleit1931} \\
F19\_045321 & NU Nor\tablenotemark{a} & 11.7 &  &--       &    &--    & LP  & L    & \citet{Meinunger1970} \\
F19\_046530 & UX Nor              & 12.1 & 2&.386(2)  &   2&.38602& CEP & CWB  & \citet{Petersen1987} \\ 
F19\_088903 & KK Nor              & 14.0 & 0&.45493(6)&    &--    & RR  & RR   & \citet{Meinunger1970} \\
F19\_089192 & EO Nor              & 11.3 & 0&.8523(3) &  0.&8523(2)& EA & EA/SD: & \citet{Kruytbosch1935}\\ 
F19\_093711 & IZ Nor              & 11.0 &  &--       &    &--    & LP  & L    & \citet{Meinunger1970} \\
F19\_098447 & NSV 7658            & 10.6 &  &--       &    &--    & LP  & --   & \citet{Luyten1938} \\
F19\_104459 & EM Nor              & 11.2 & 0&.7383(2) &  0&.7384  & EW  & EW   & \citet{Malkov2006} \\
F19\_107786 & IX Nor              & 10.6 &  &--       &    &--    & LP  & M    & \citet{Meinunger1970} \\
F19\_111712 & UV Nor              & 12.8 & 0&.8742(2) &  0&.8741  & EA  & EA   & \citet{Malkov2006} \\
F19\_116322 & PW Nor              & 12.1 &  &--       &    &--    & LP  & M:   & \citet{Luyten1936} \\
\tableline
\end{tabular}
\tablenotetext{a}{\scriptsize\citet{Meinunger1970} assigns the variability of NU~Nor to the star 2MASS\,16274939-5533450 (BEST\,II F19\_045485). However, in BEST\,II data, the variability can clearly be assigned to the object 2MASS\,16275033-5533400 (BEST\,II F19\_045321), which is located at an angular distance of $9''$ to the former.}
\tablecomments{Given are identifiers of this work and the GCVS and/or VSX, BEST\,II instrumental magnitudes~$R_B$, and periods and classifications (if available) as obtained within this work and by previous surveys as referenced, respectively.}
\end{table*}

A figure set of all light curves is available in the online version of the journal. Figure~\ref{fig:lcs} illustrates its form and content by highlighting examples that can be interesting for further astrophysical studies, such as, e.g., eclipsing binaries with a high SNR (F17\_10421, F19\_009645, F19\_030794, F19\_033571, F19\_100160), eccentric eclipsing binaries (F19\_055270, F19\_100956), a cataclysmic binary (F19\_022713), or RR Lyrae pulsators with amplitude modulation known as the Blazhko-effect (\citealt{Blazhko1907}; F19\_086712, F19\_124221).

\subsection{Known Variables}\label{sec:knownvars}
\begin{figure*}[p]
\begin{tabular}{@{}p{0.23\textwidth}@{}p{0.23\textwidth}@{}p{0.23\textwidth}@{}p{0.23\textwidth}@{}}
\includegraphics[width=\linewidth]{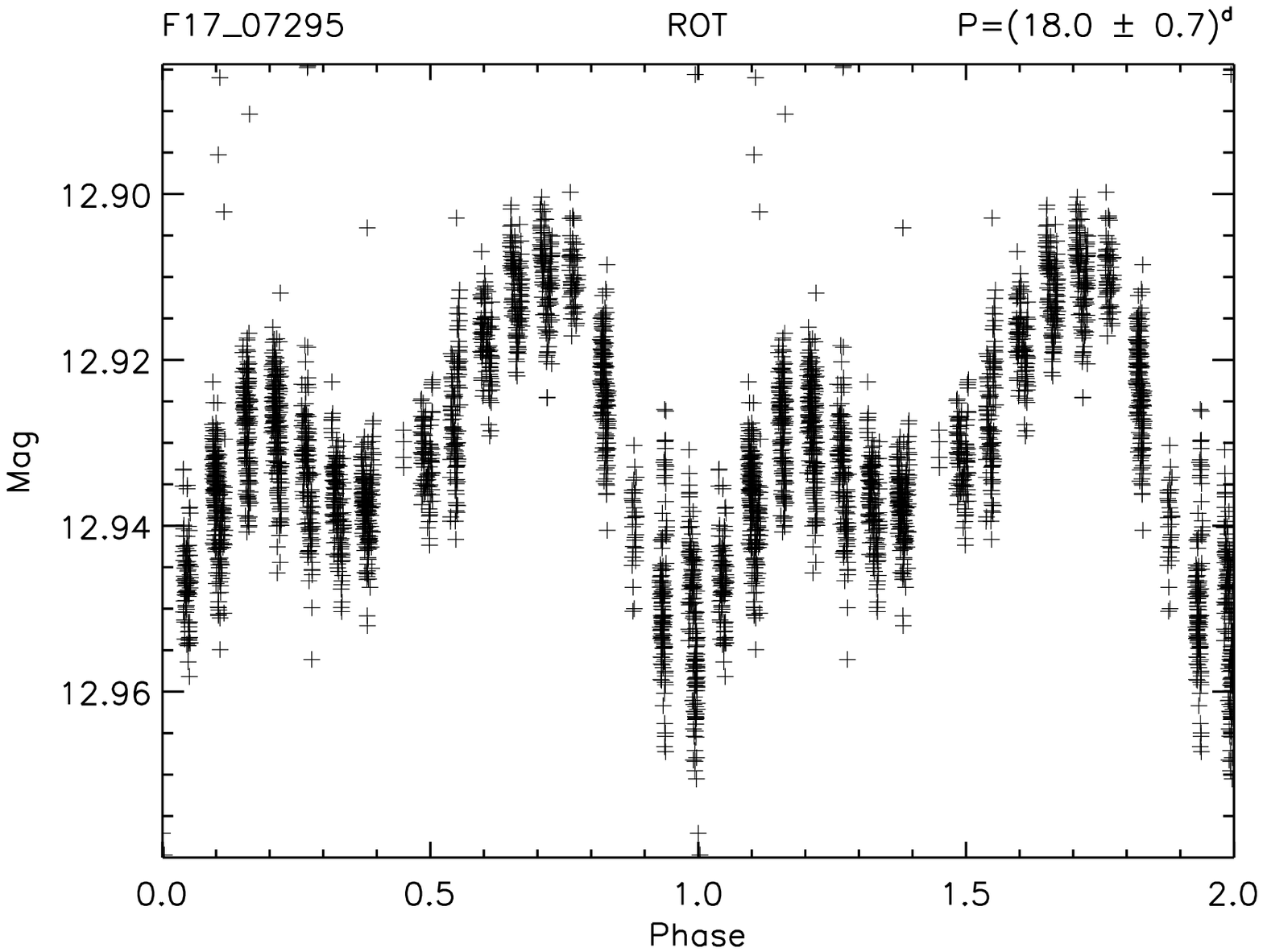} &
\includegraphics[width=\linewidth]{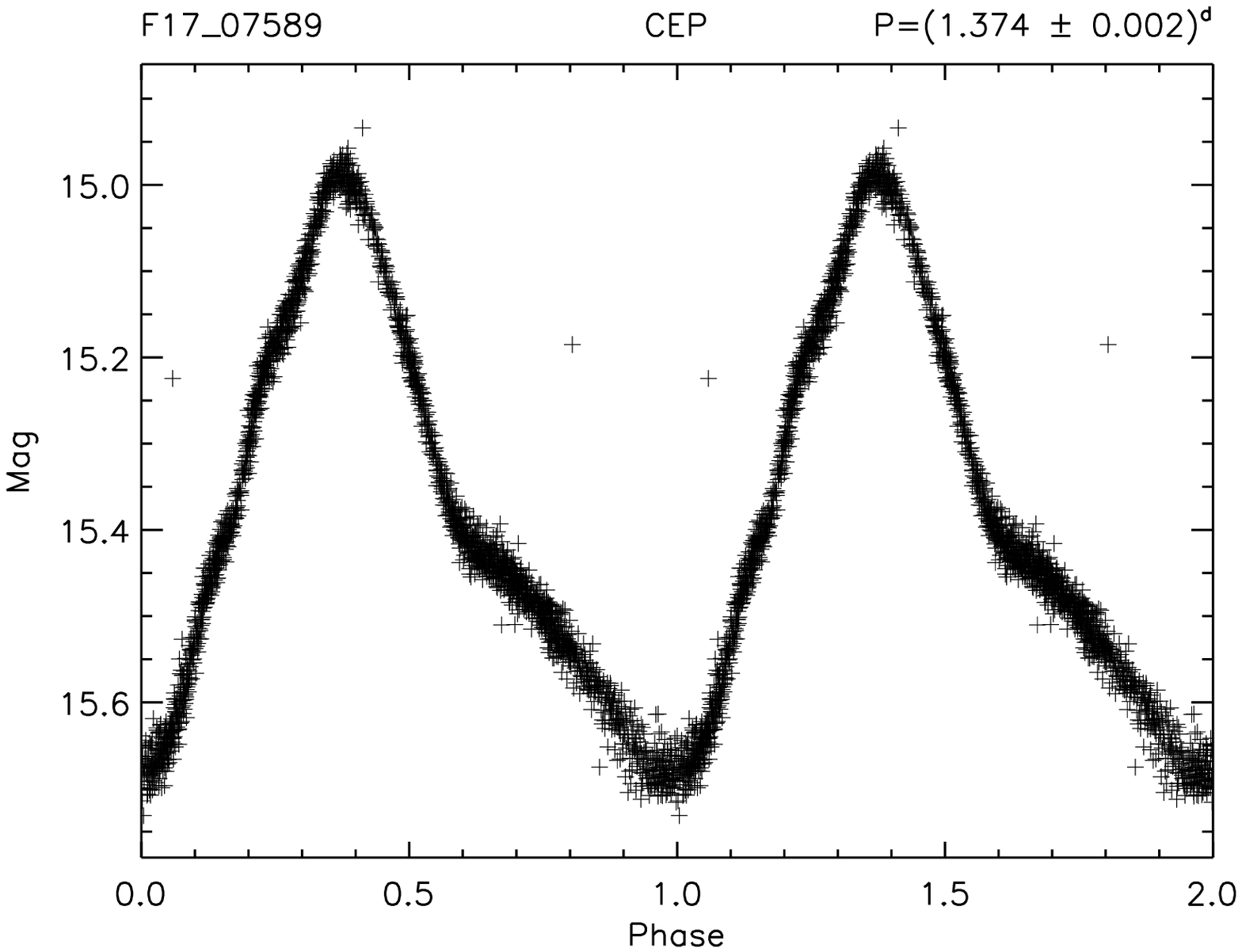} &
\includegraphics[width=\linewidth]{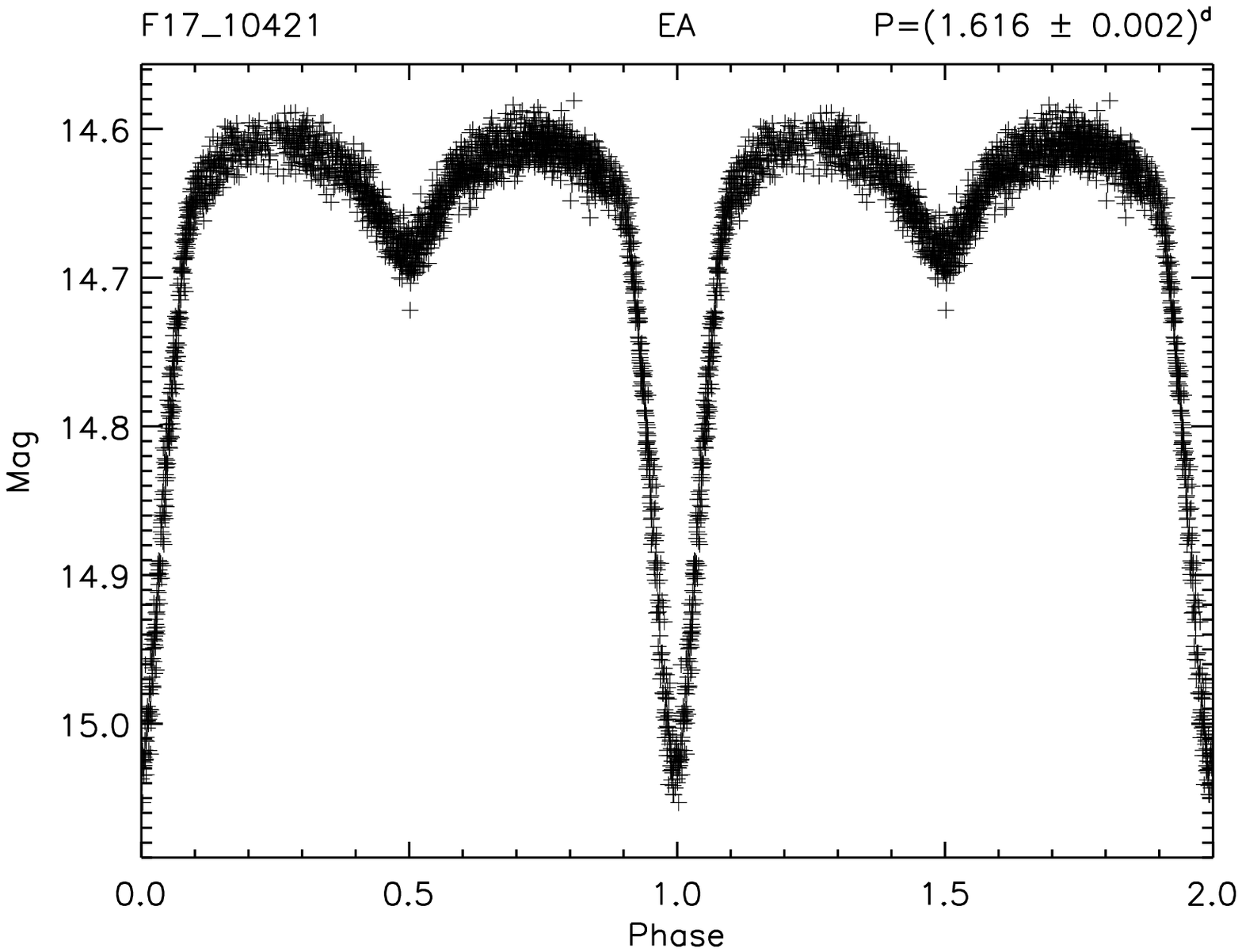} &
\includegraphics[width=\linewidth]{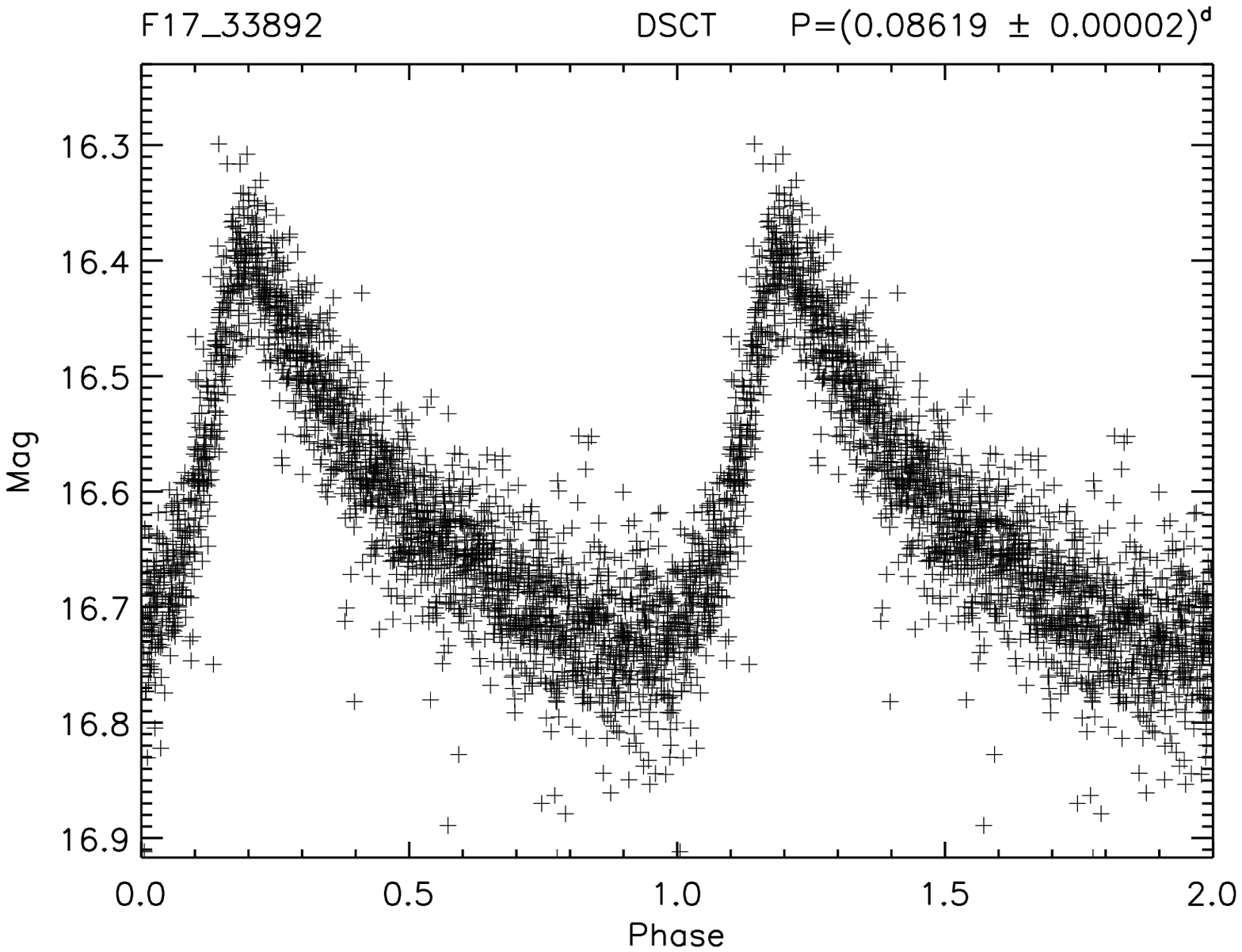} \\
\includegraphics[width=\linewidth]{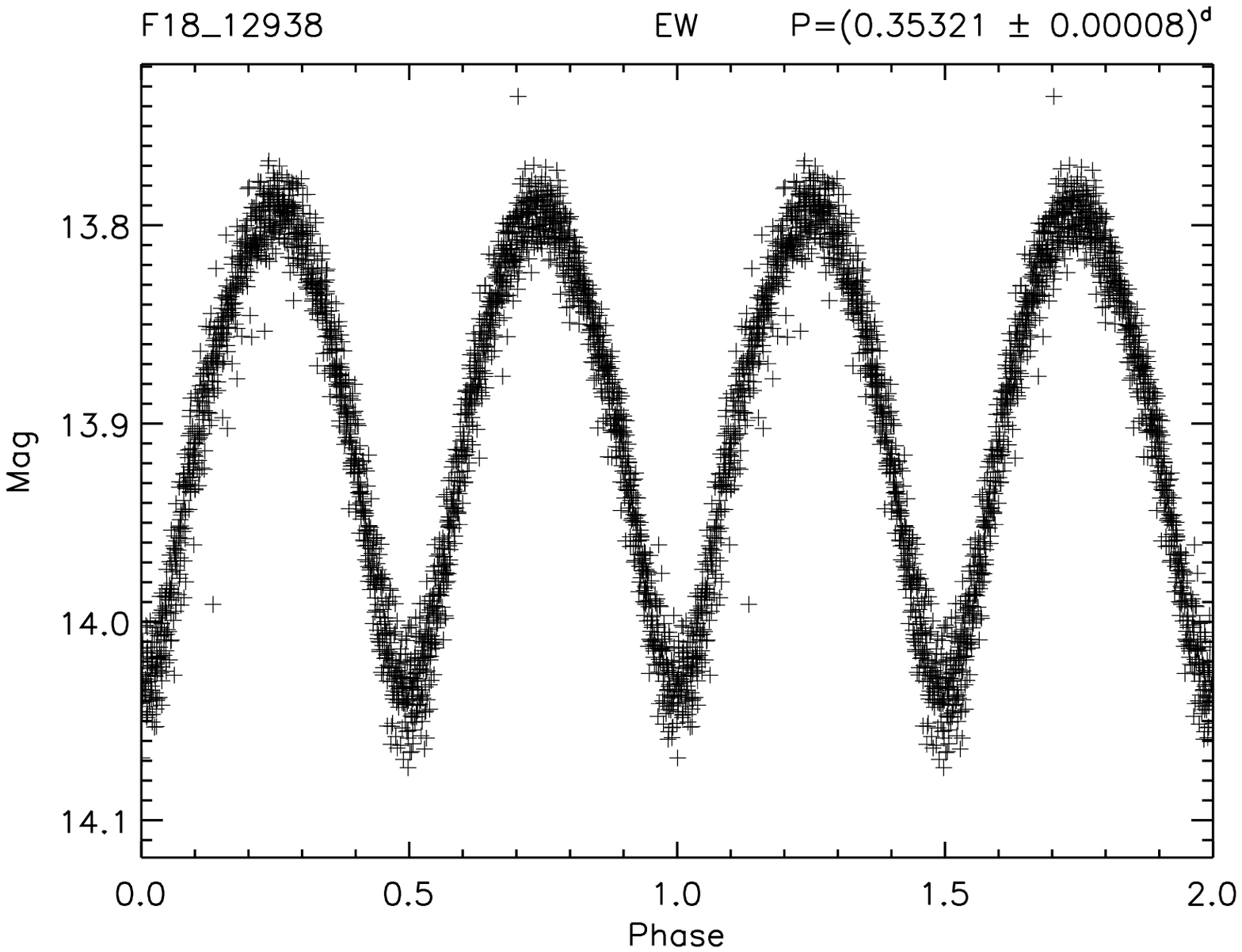} &
\includegraphics[width=\linewidth]{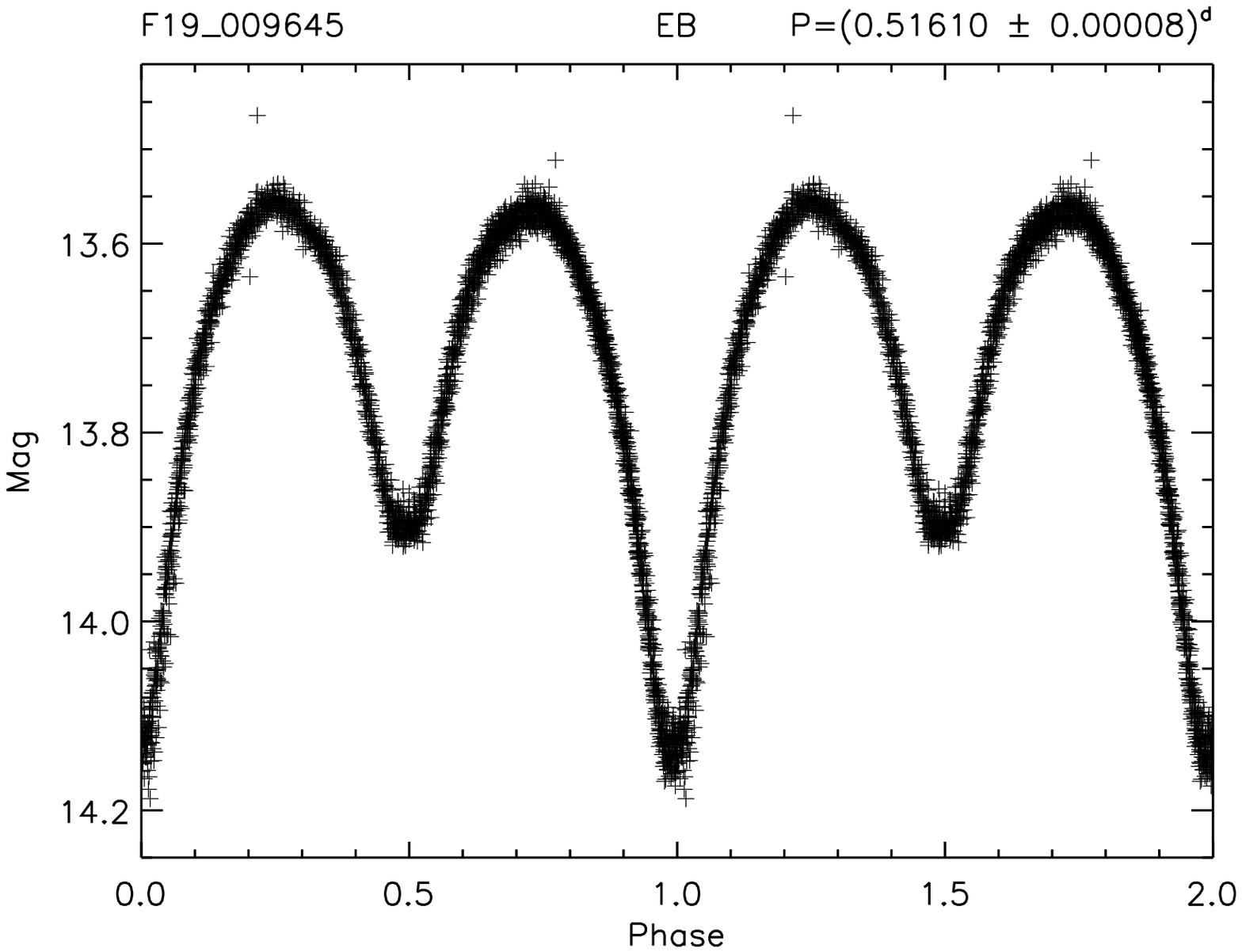} &
\includegraphics[width=\linewidth]{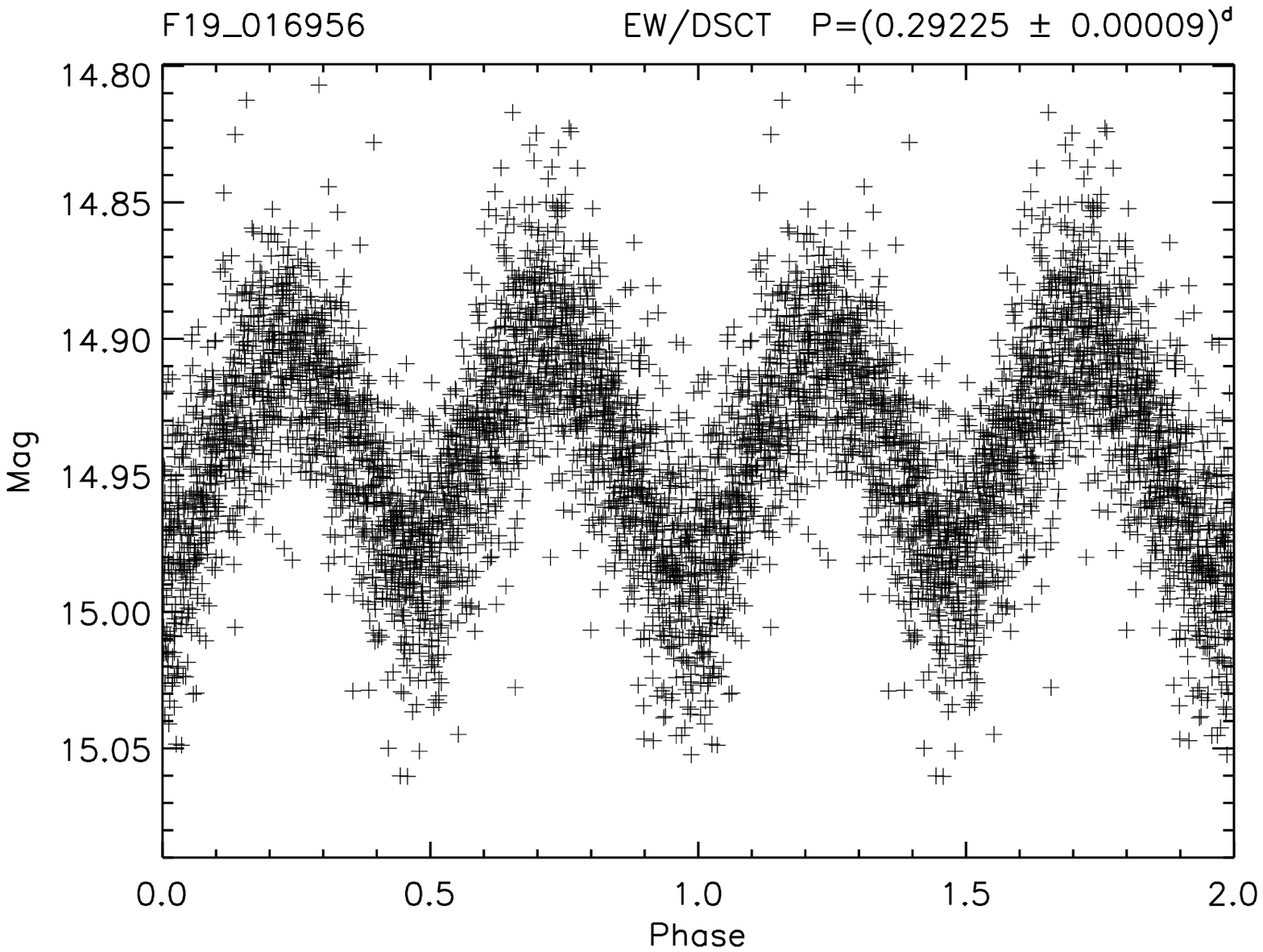} &
\includegraphics[width=\linewidth]{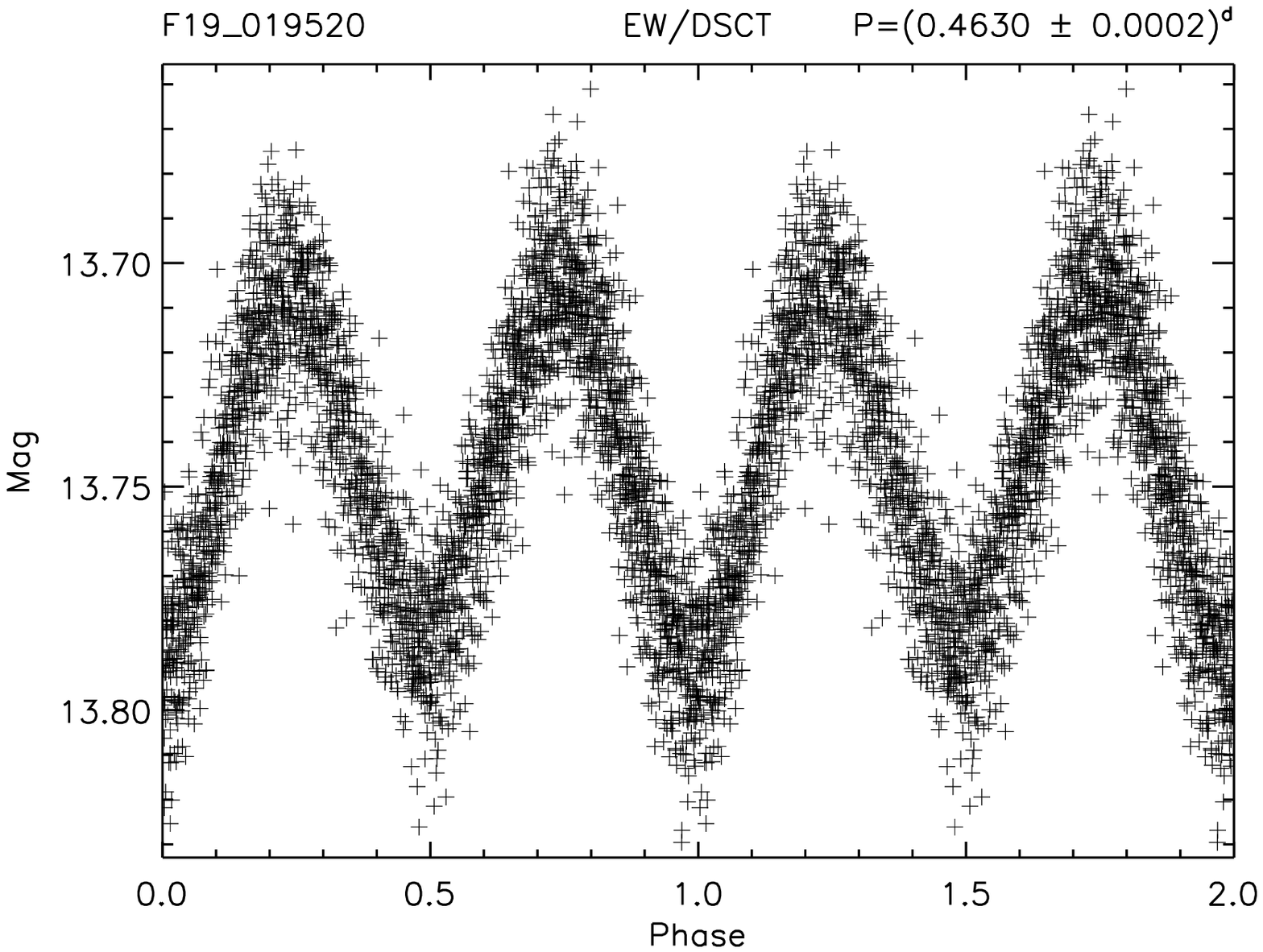} \\
\includegraphics[width=\linewidth]{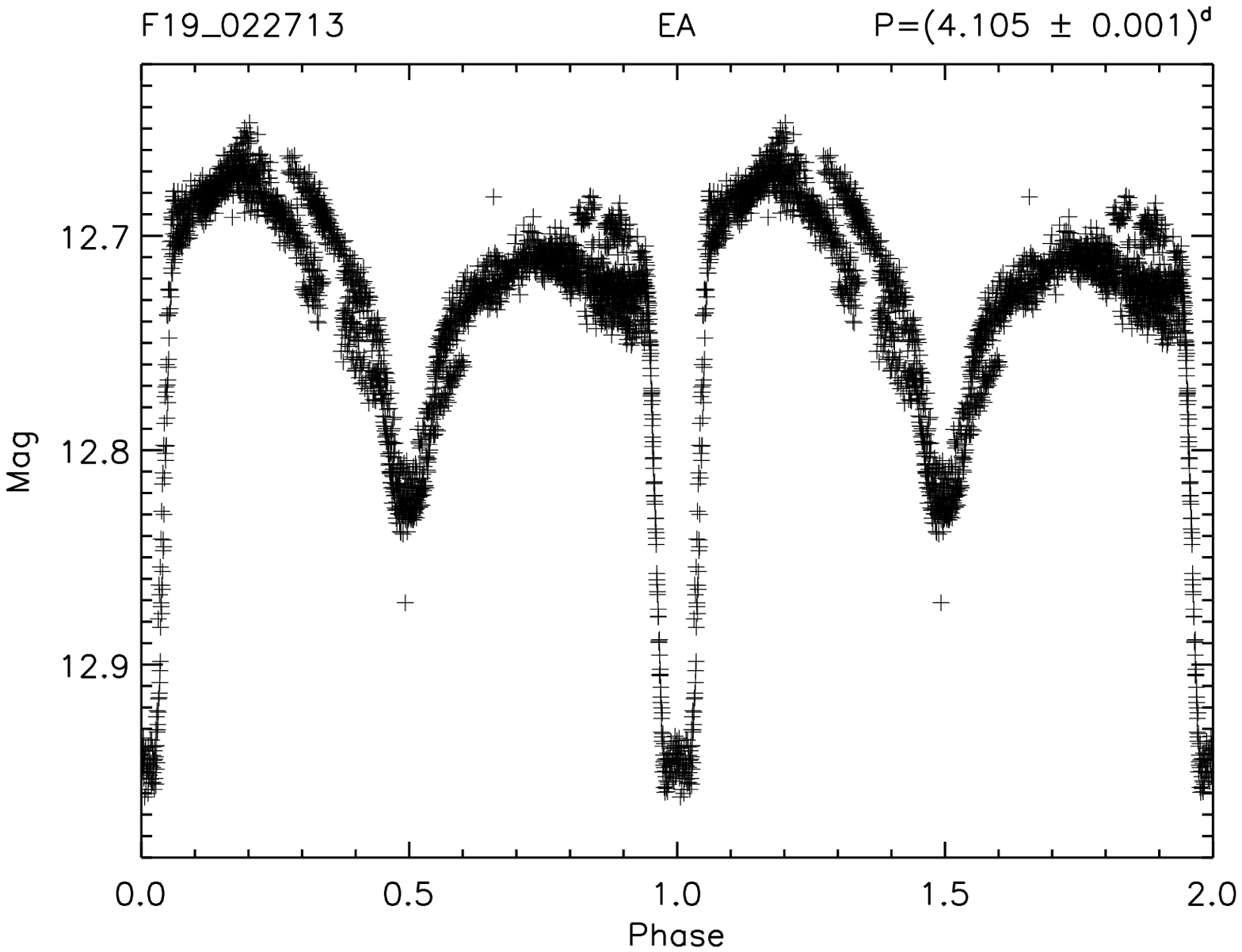} &
\includegraphics[width=\linewidth]{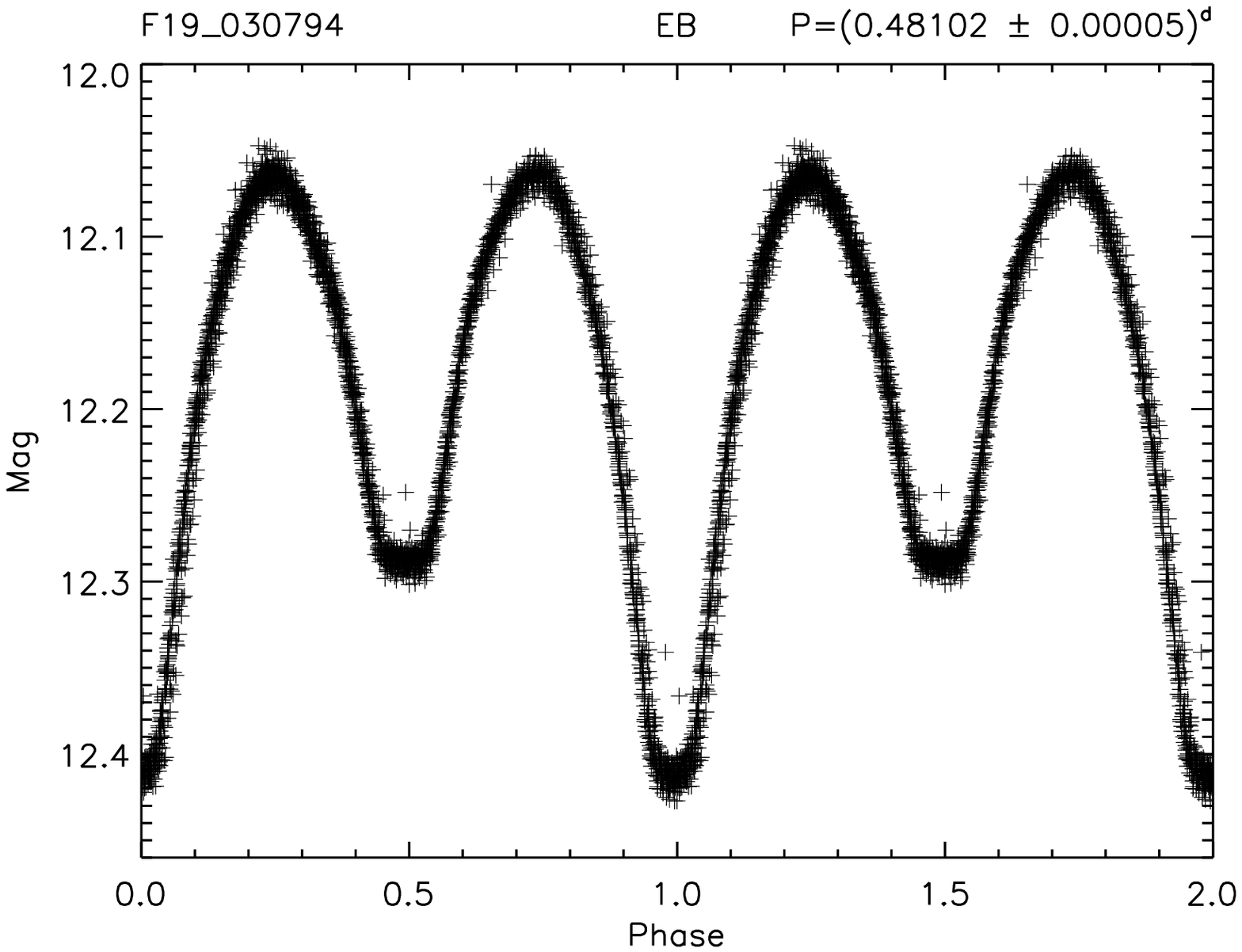} &
\includegraphics[width=\linewidth]{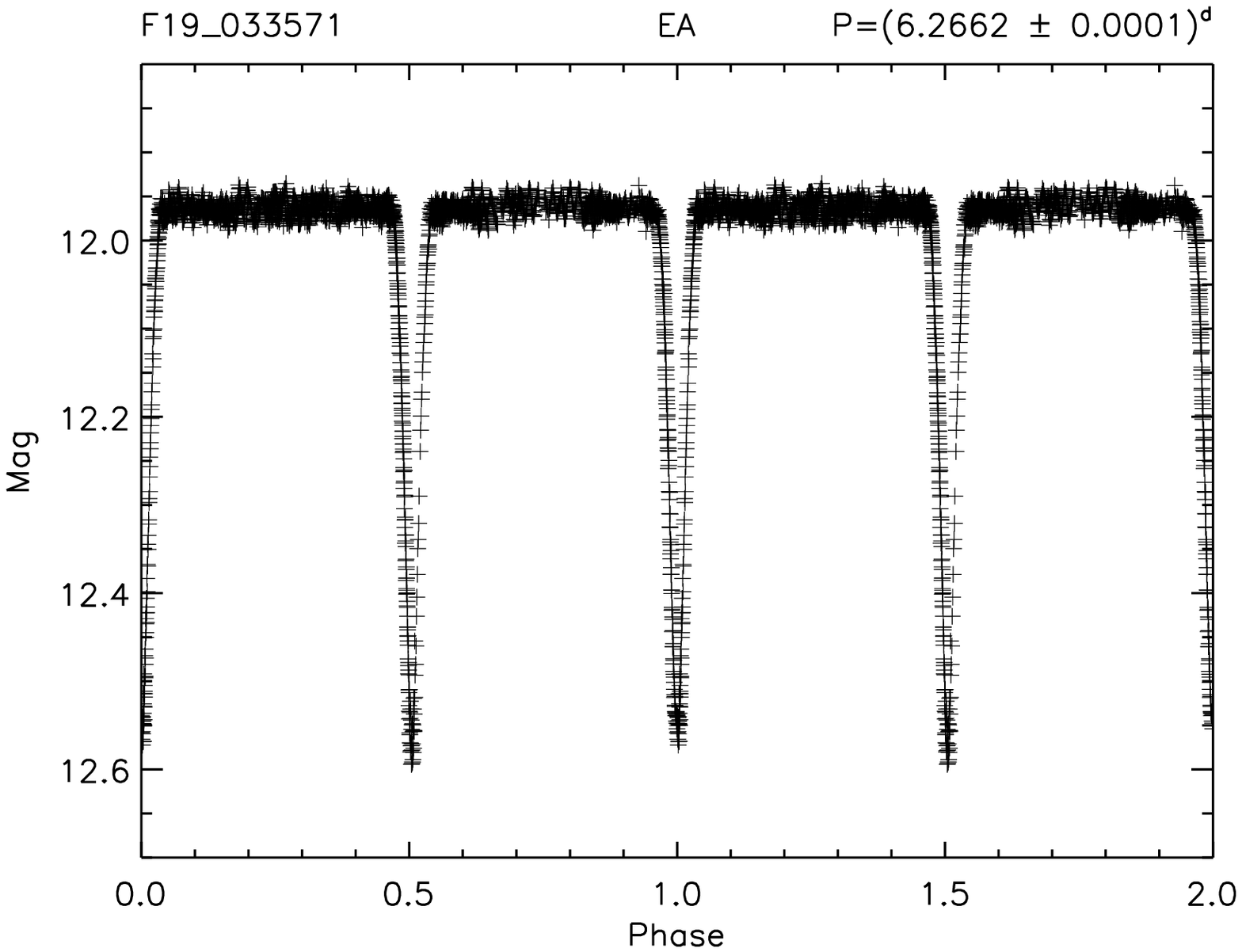} &
\includegraphics[width=\linewidth]{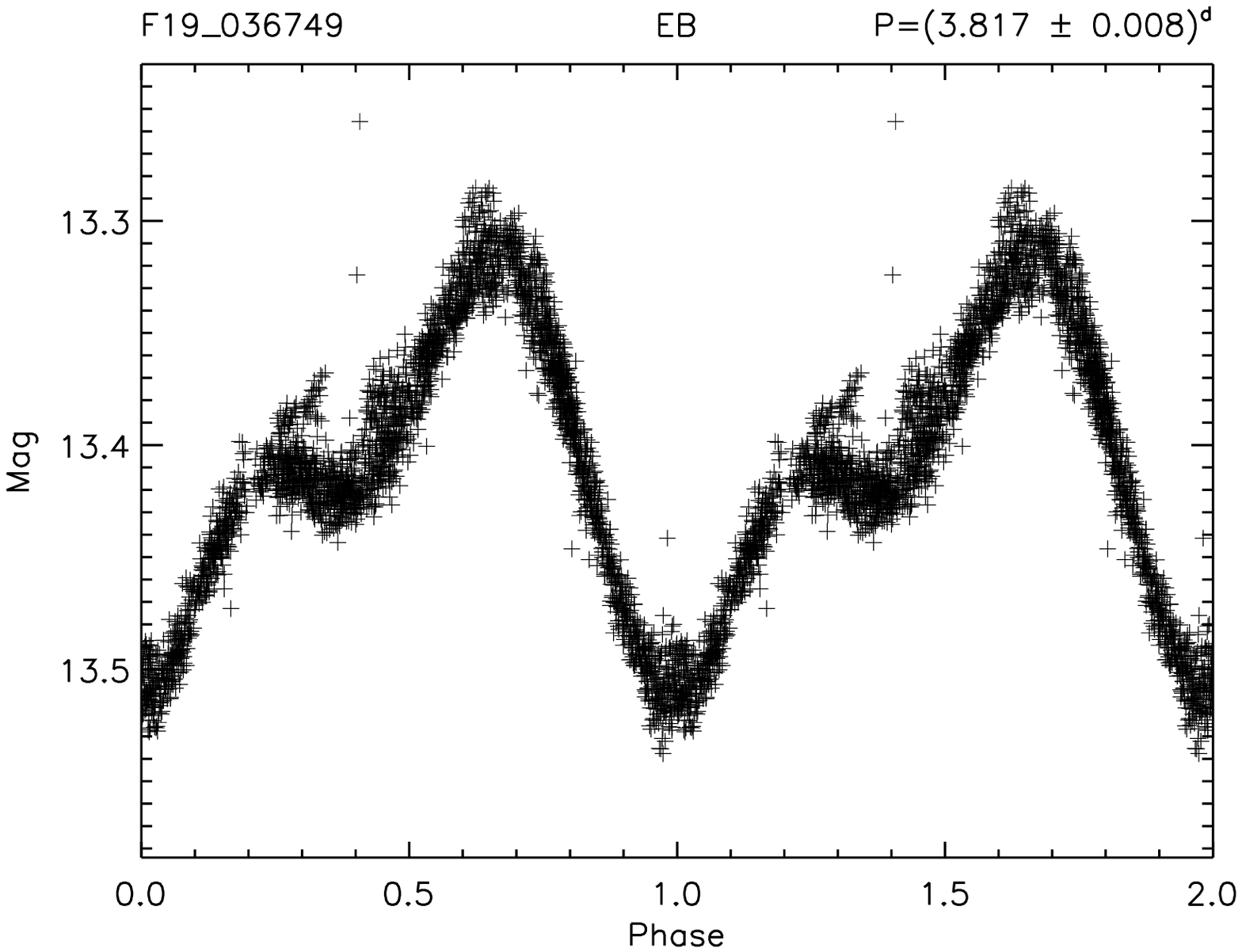} \\
\includegraphics[width=\linewidth]{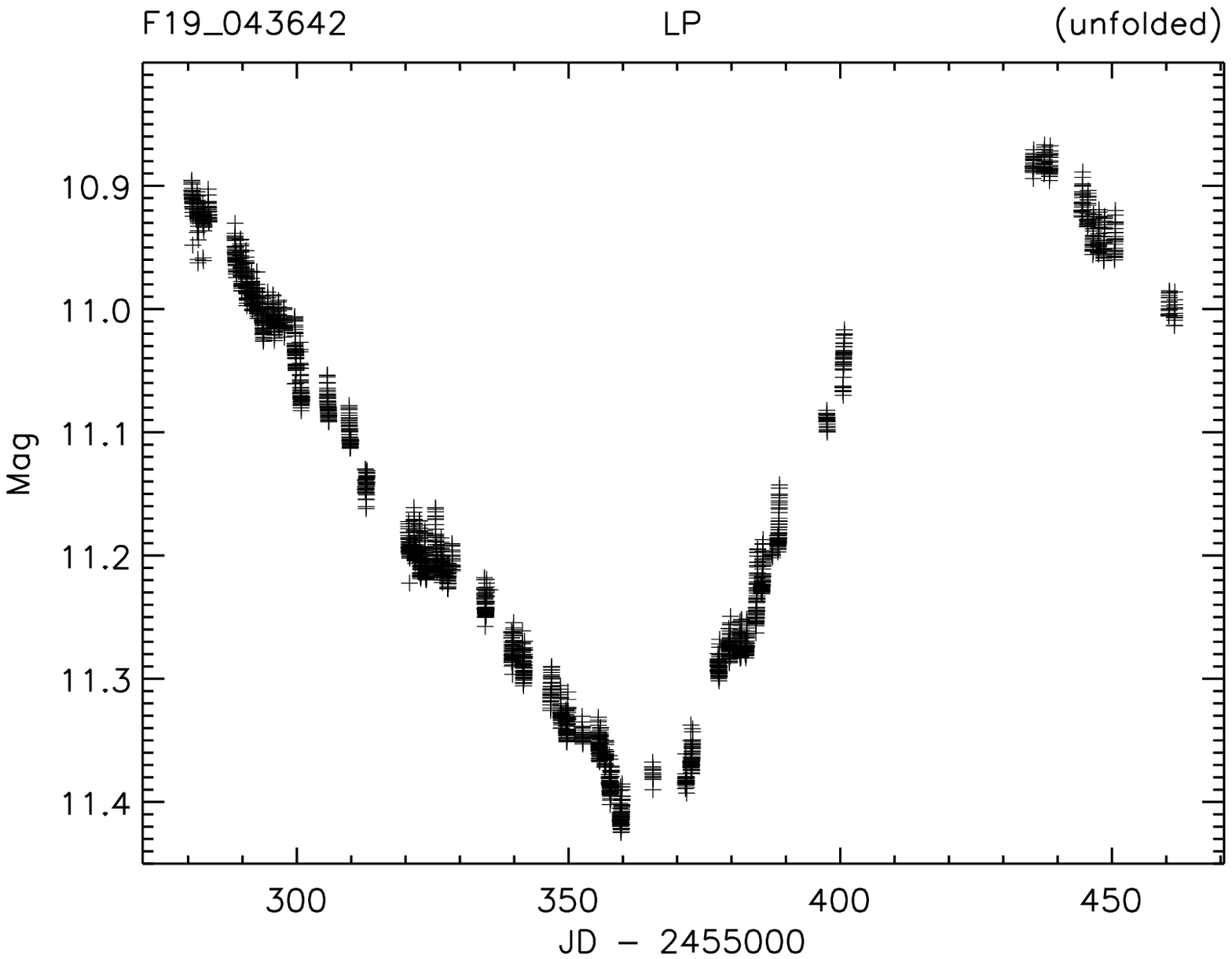} &
\includegraphics[width=\linewidth]{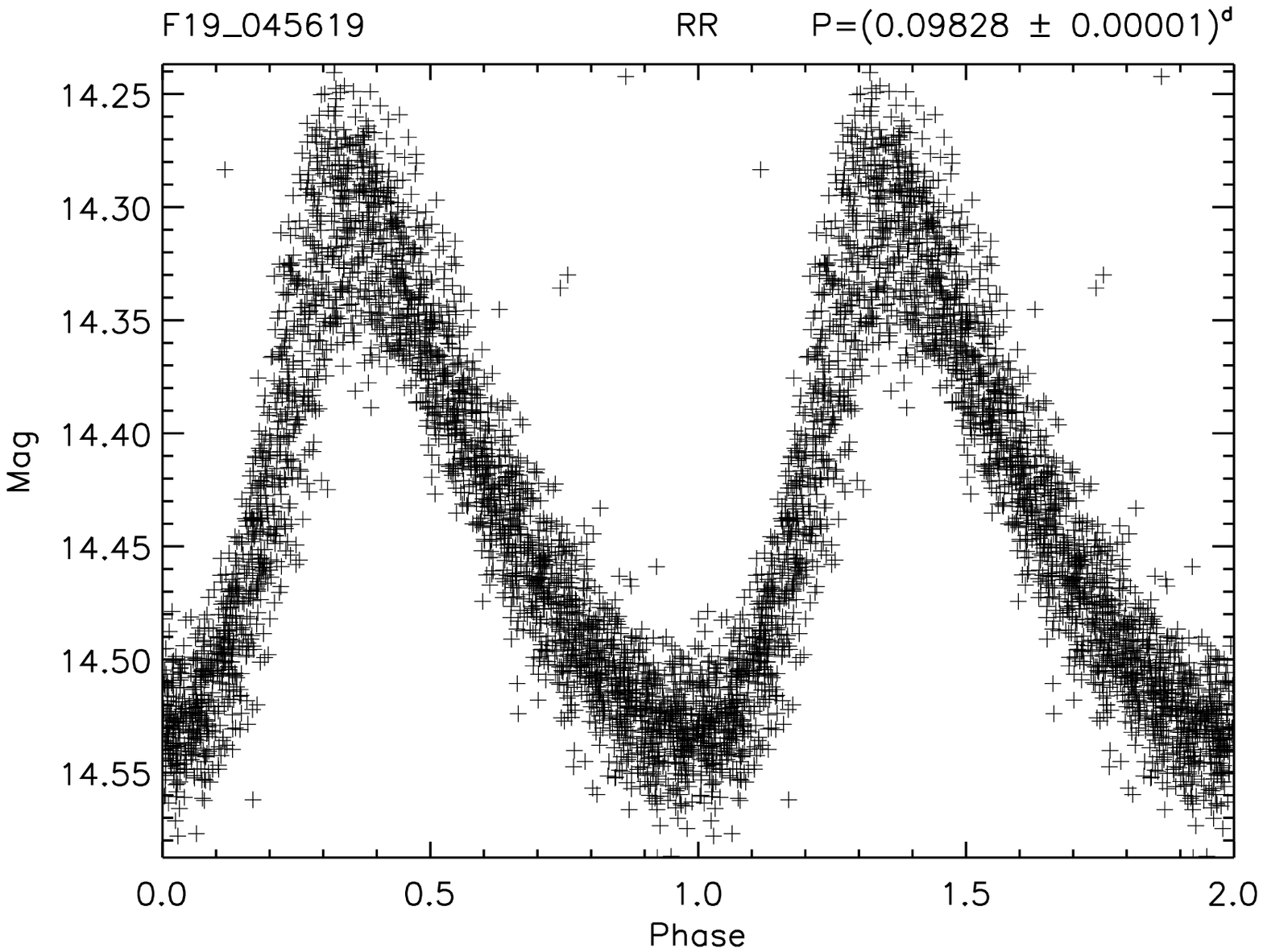} &
\includegraphics[width=\linewidth]{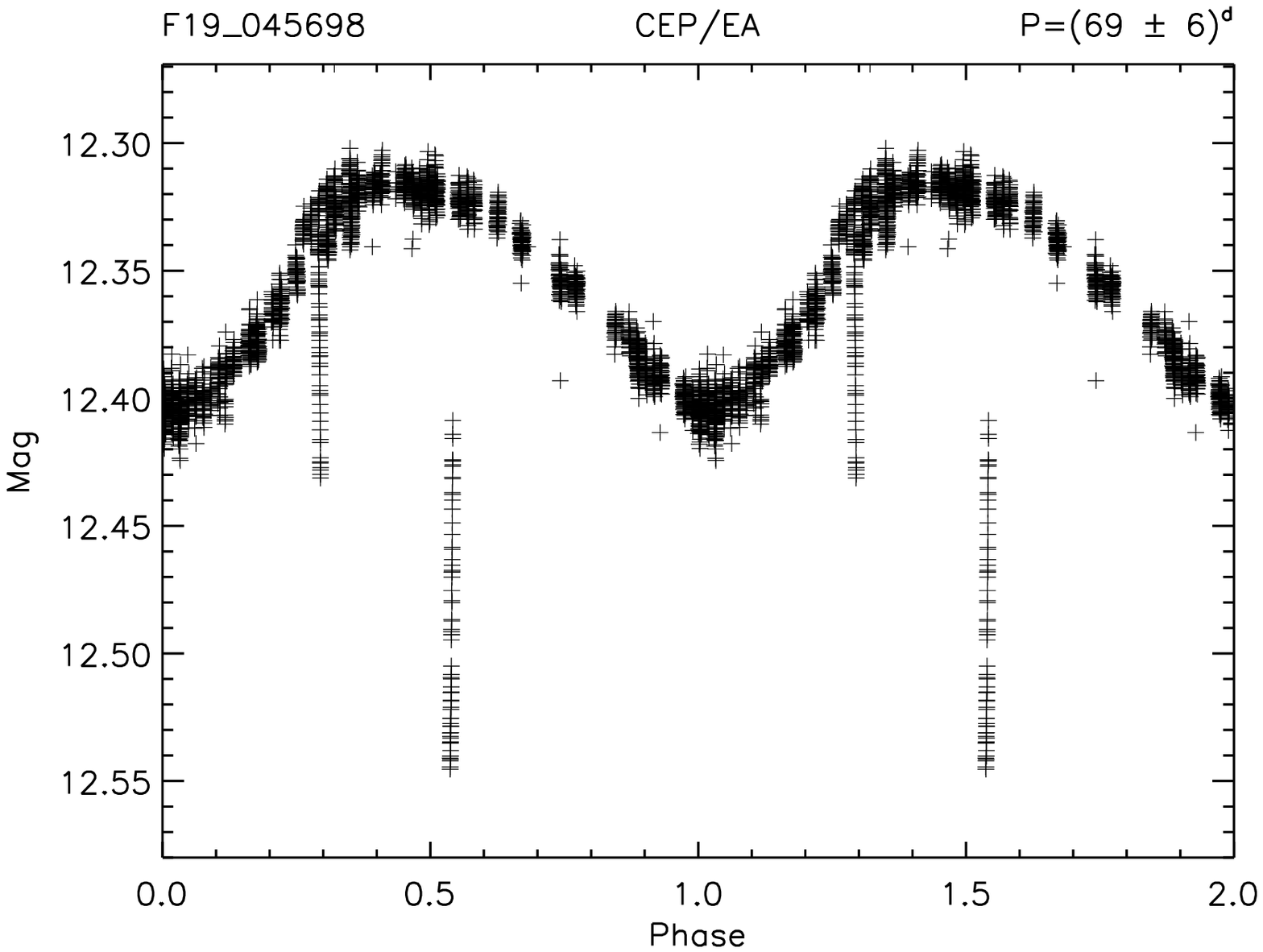} &
\includegraphics[width=\linewidth]{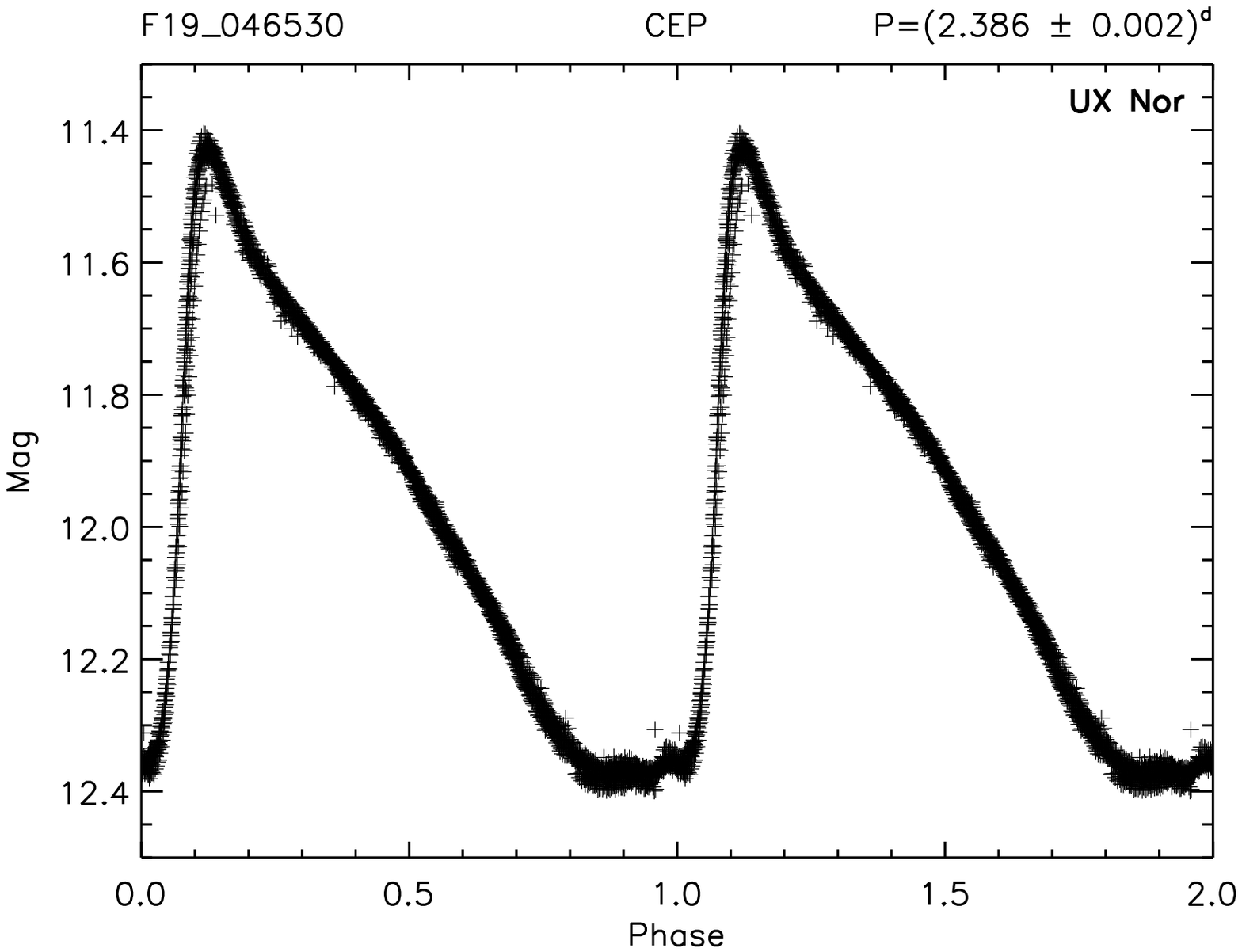} \\
\includegraphics[width=\linewidth]{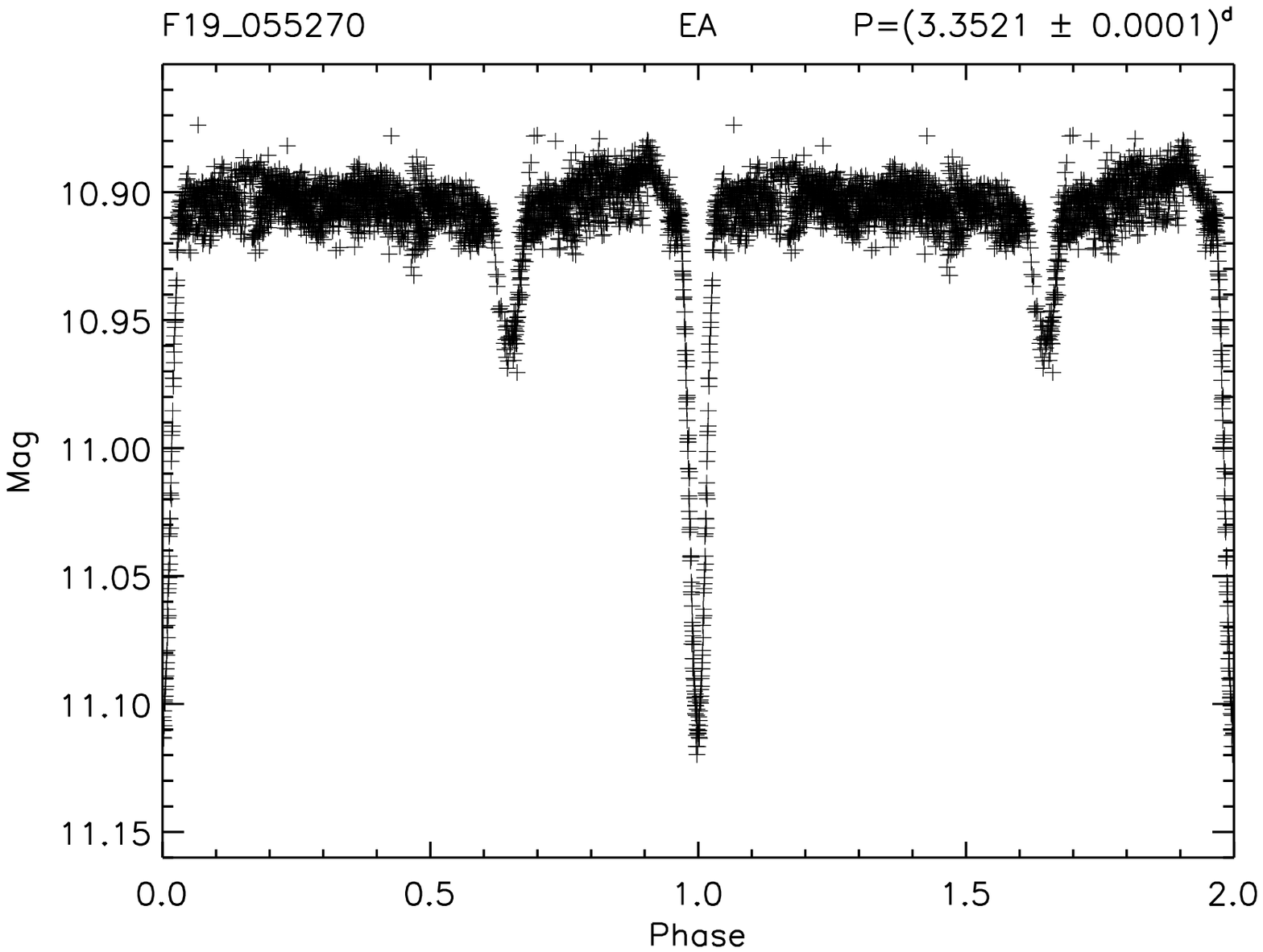} &
\includegraphics[width=\linewidth]{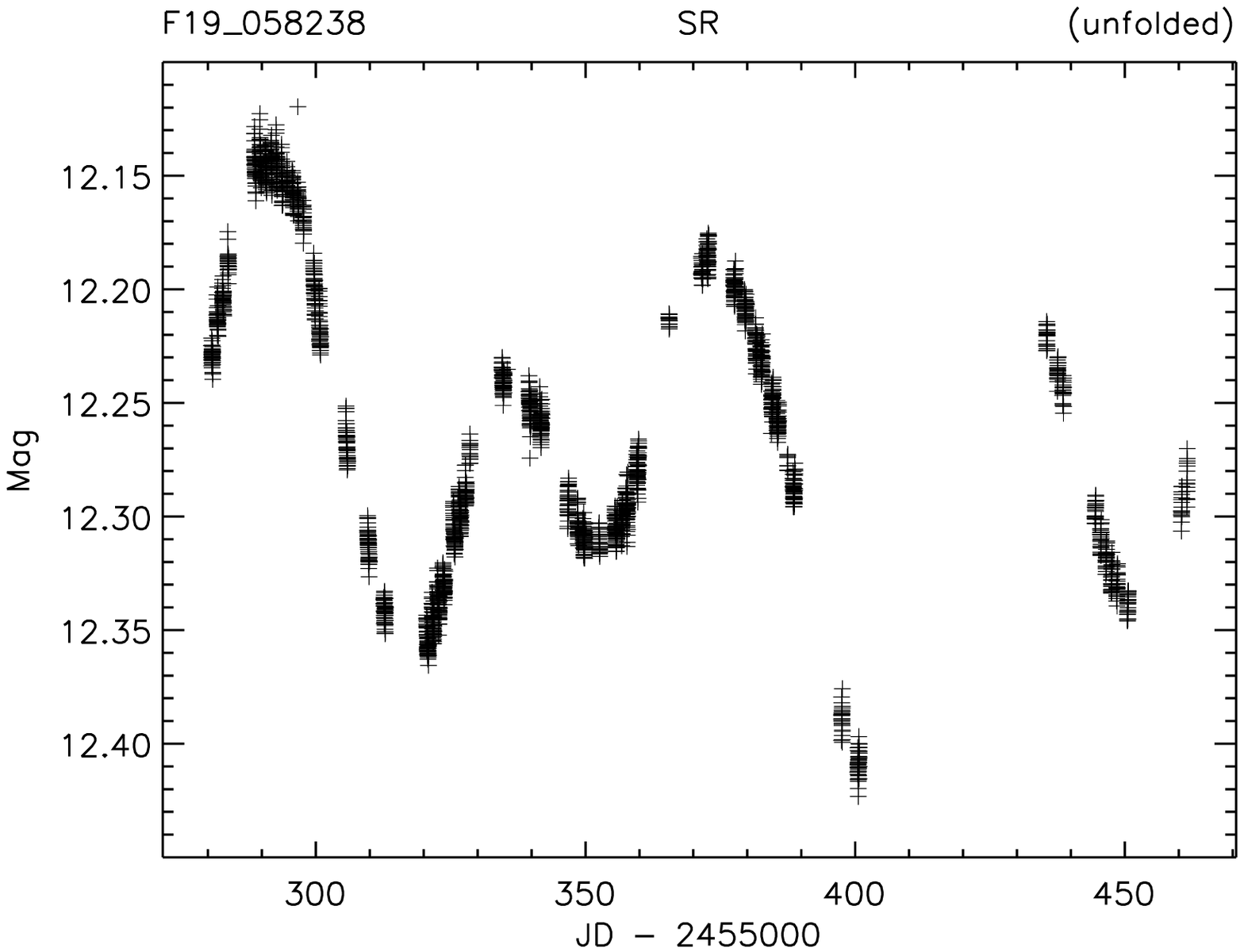} &
\includegraphics[width=\linewidth]{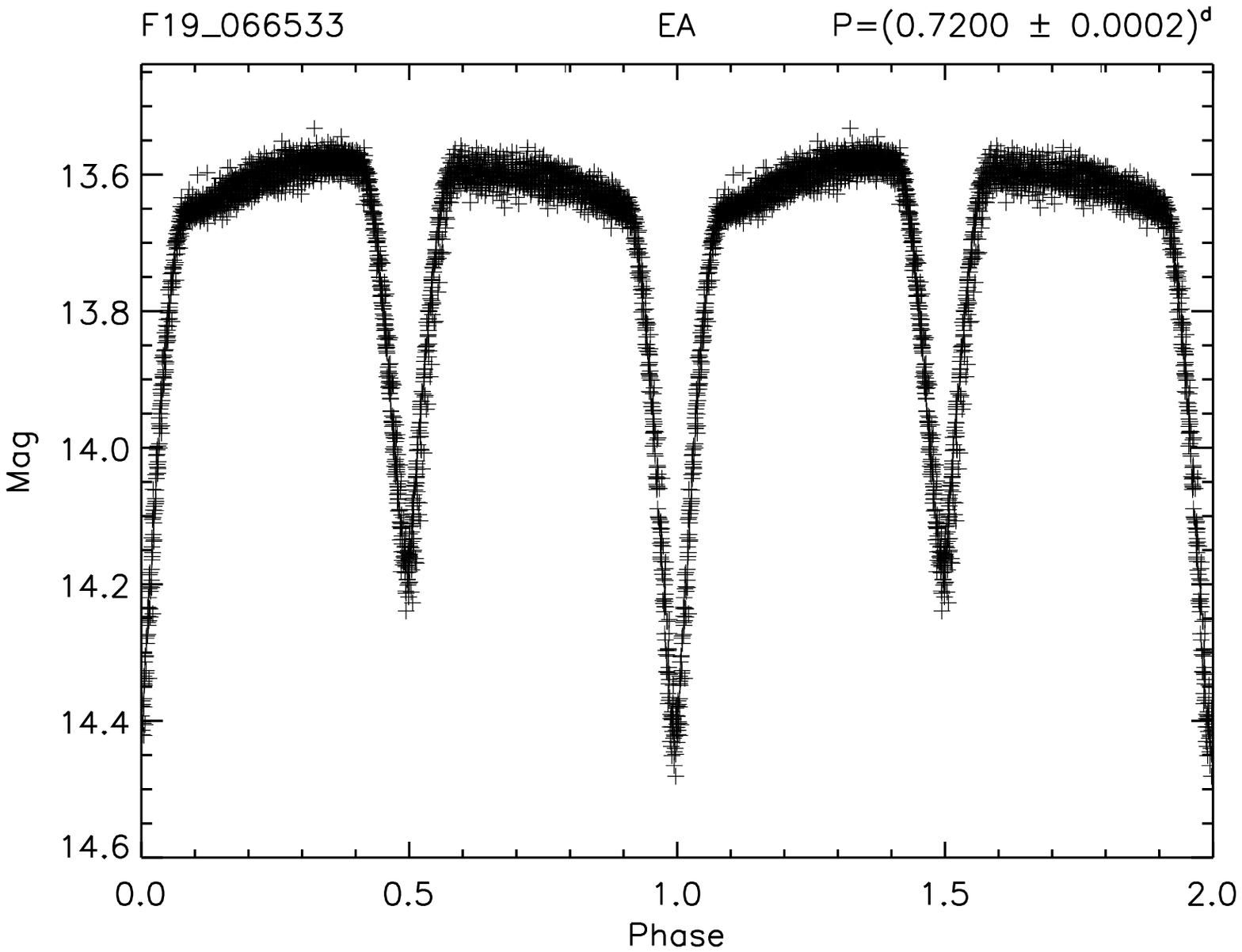} &
\includegraphics[width=\linewidth]{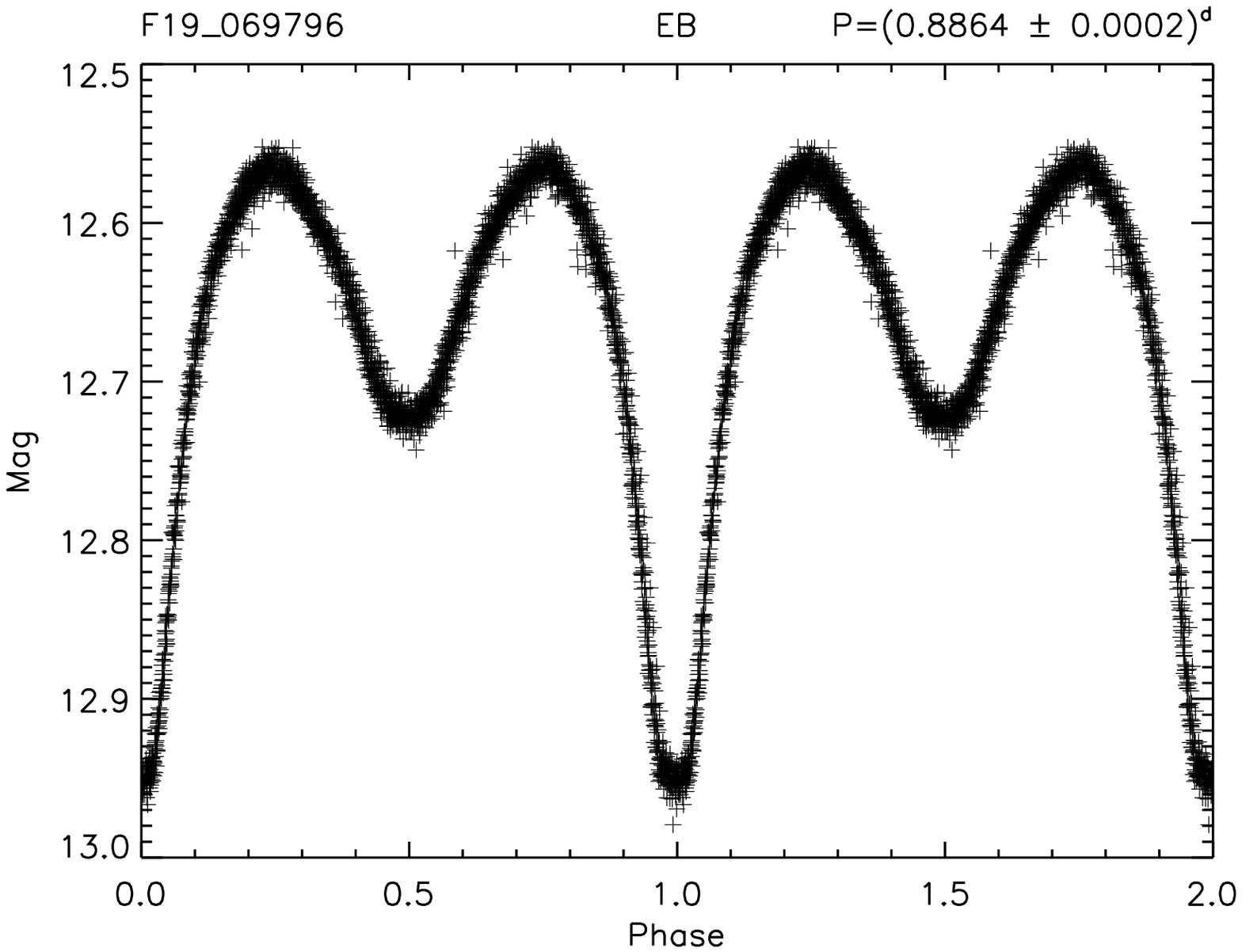} \\
\includegraphics[width=\linewidth]{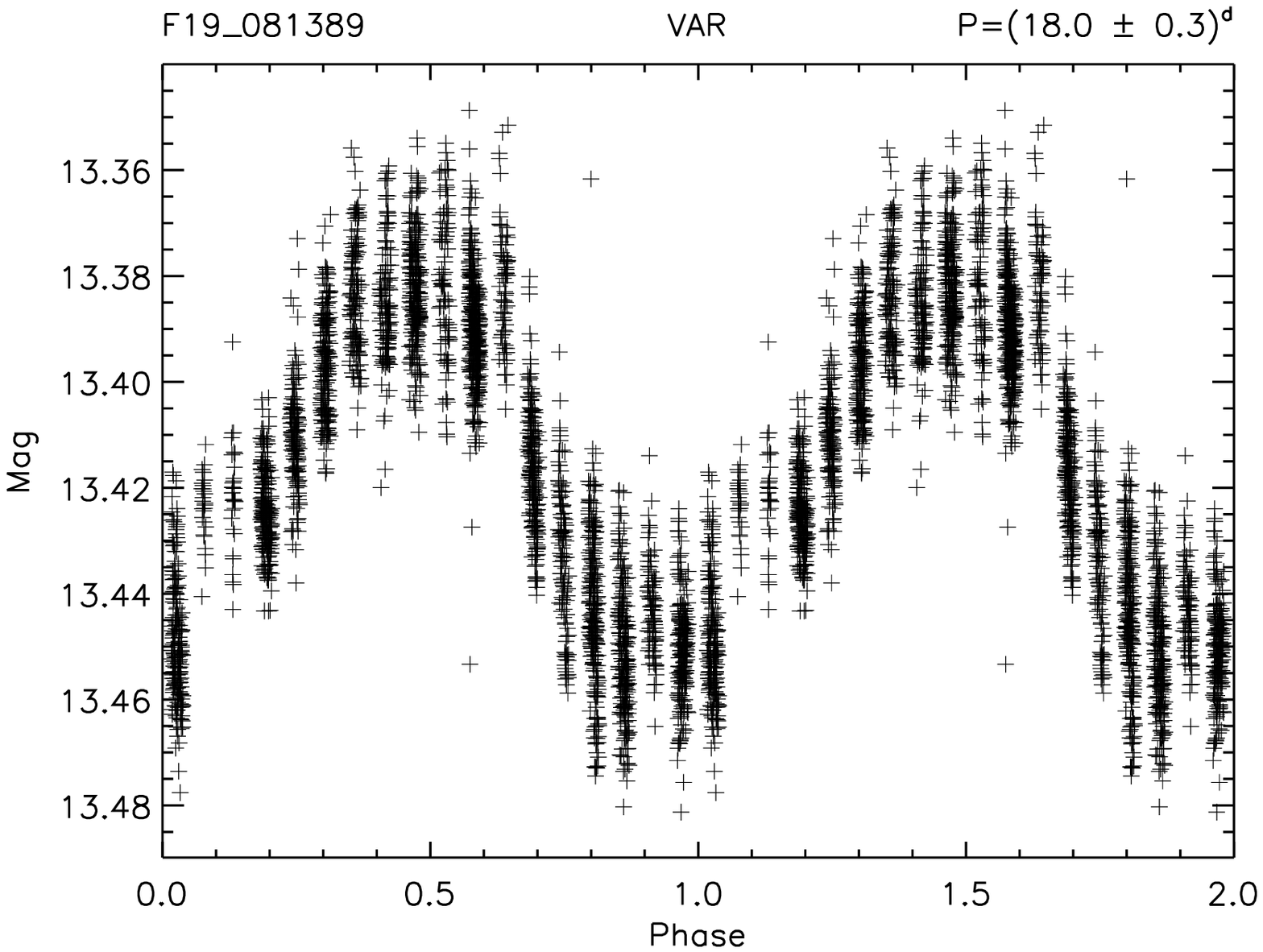} &
\includegraphics[width=\linewidth]{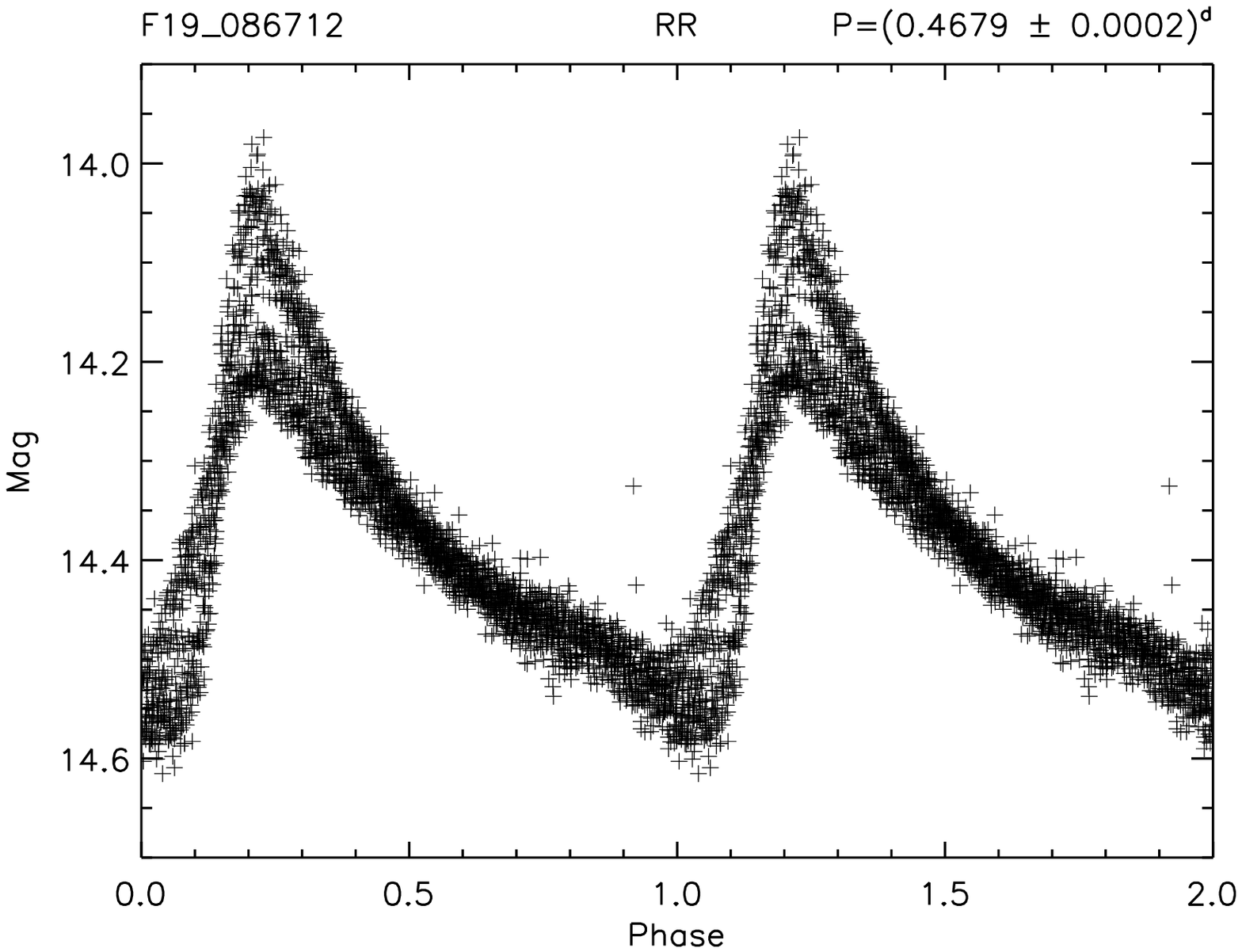} &
\includegraphics[width=\linewidth]{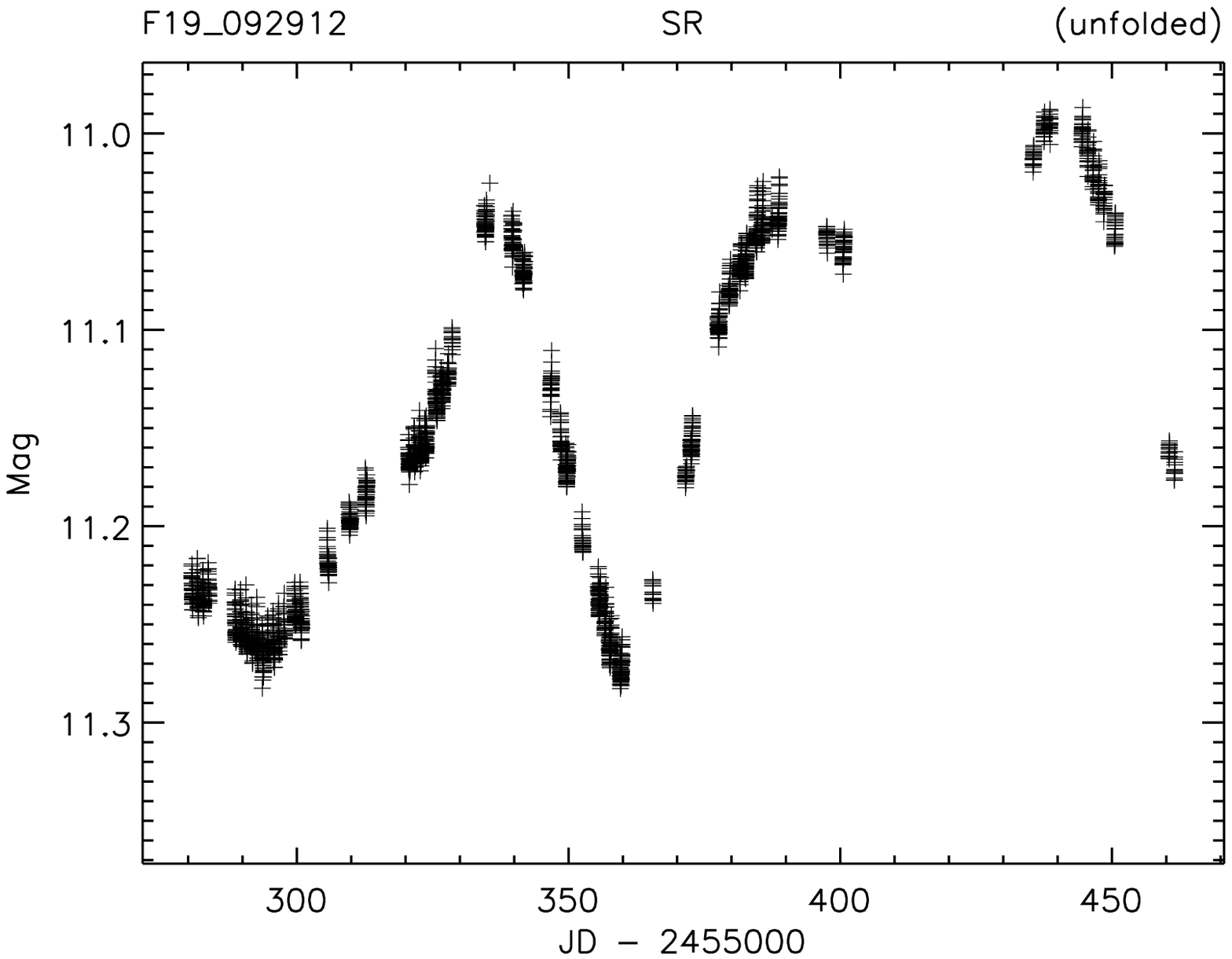} &
\includegraphics[width=\linewidth]{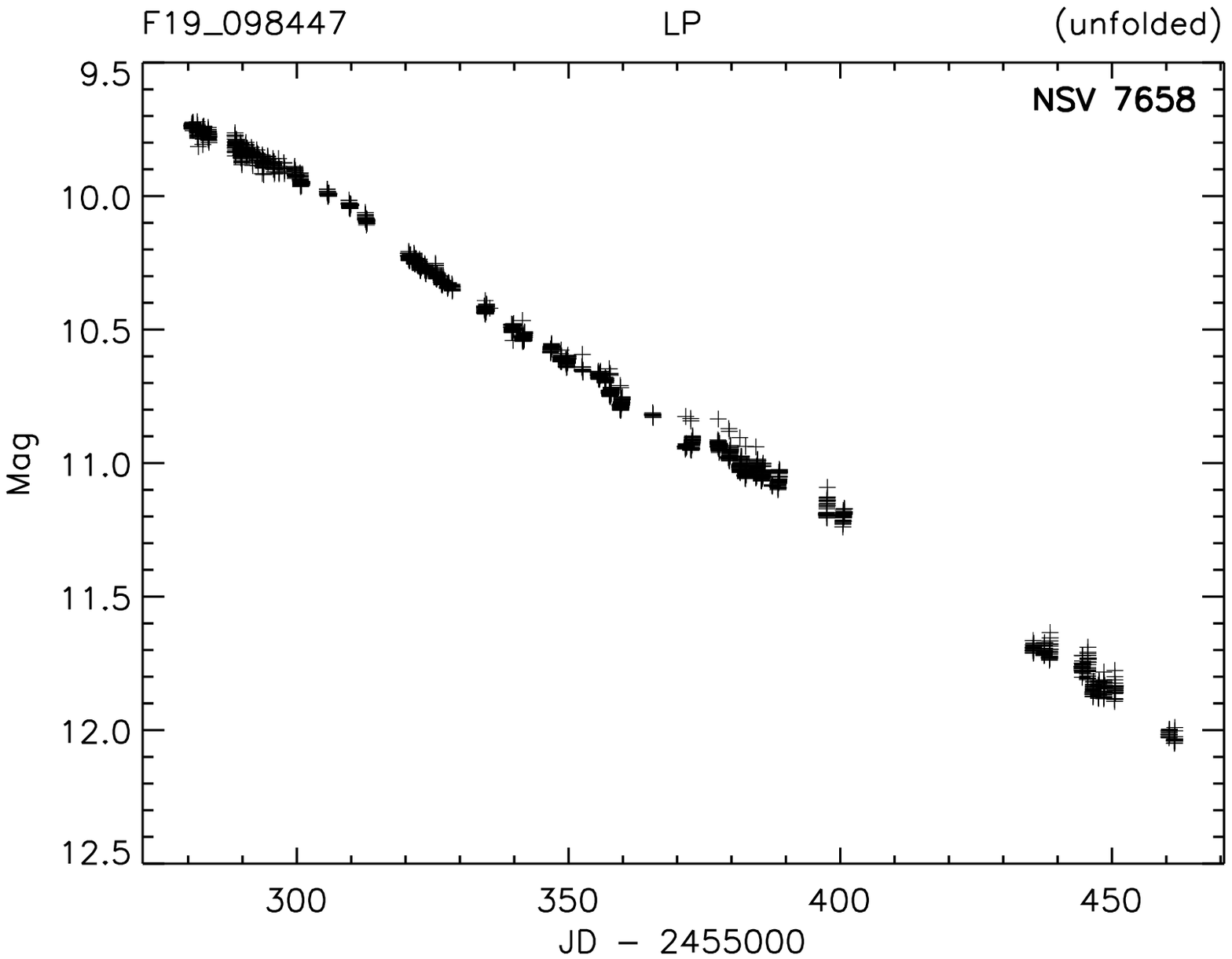} \\
\includegraphics[width=\linewidth]{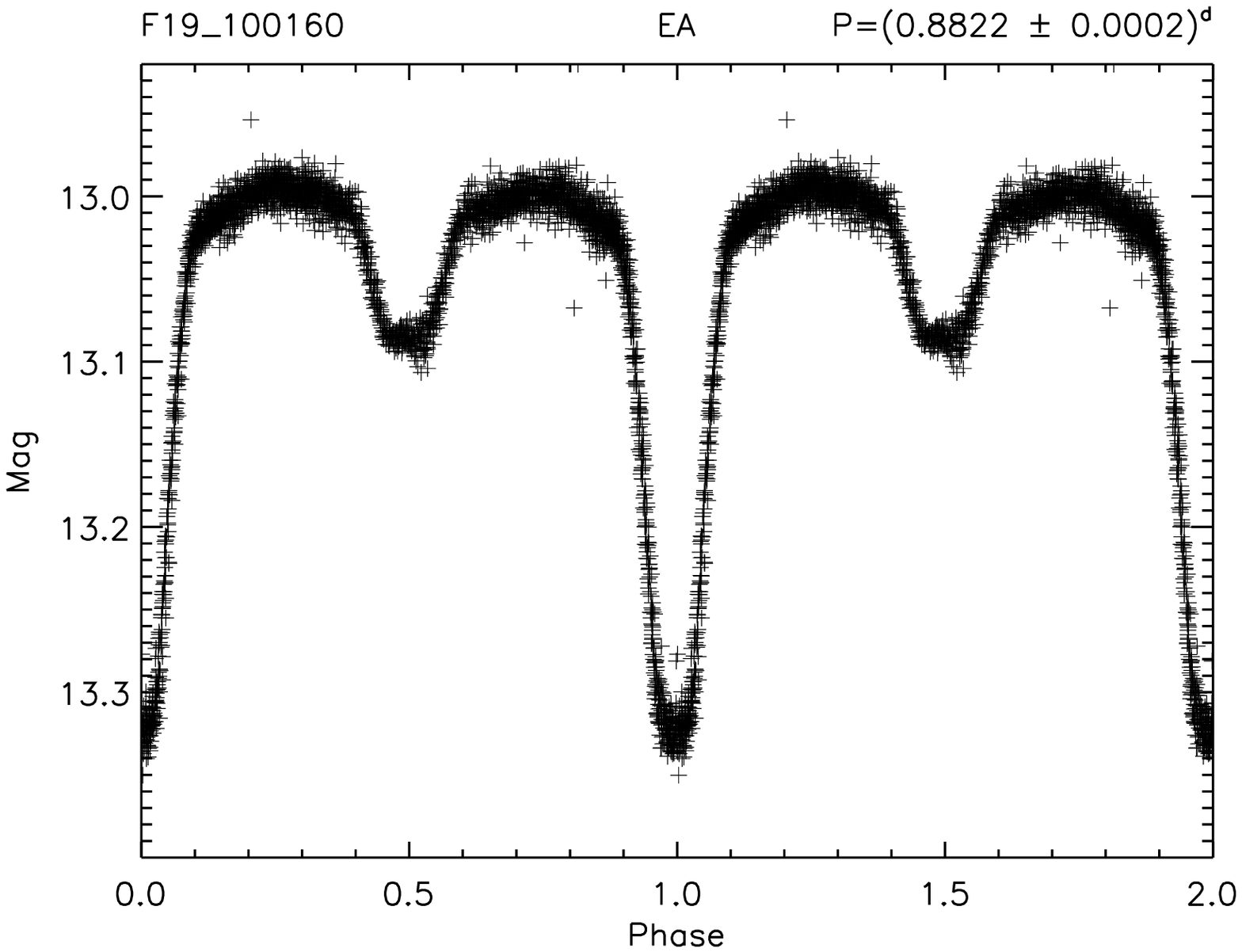} &
\includegraphics[width=\linewidth]{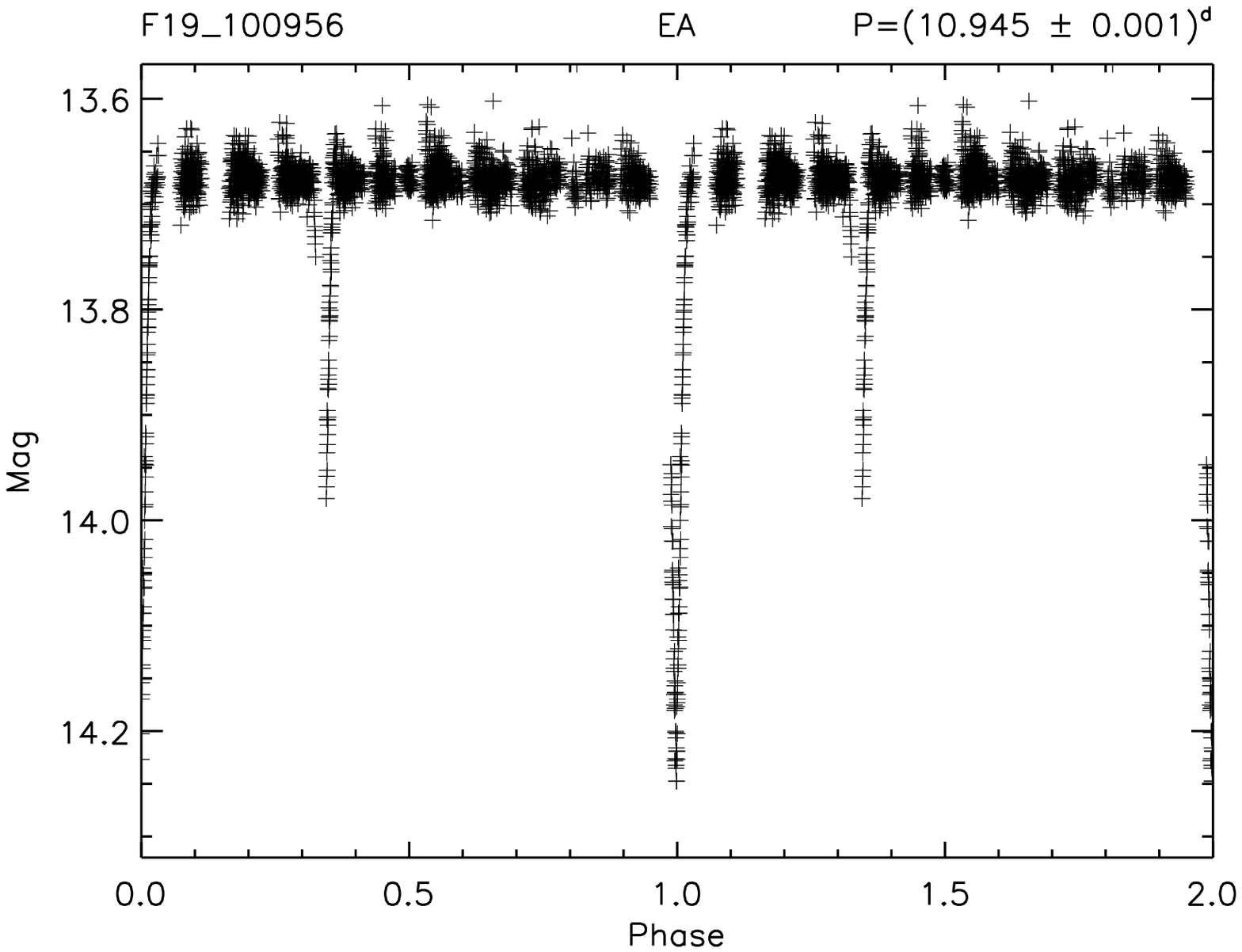} &
\includegraphics[width=\linewidth]{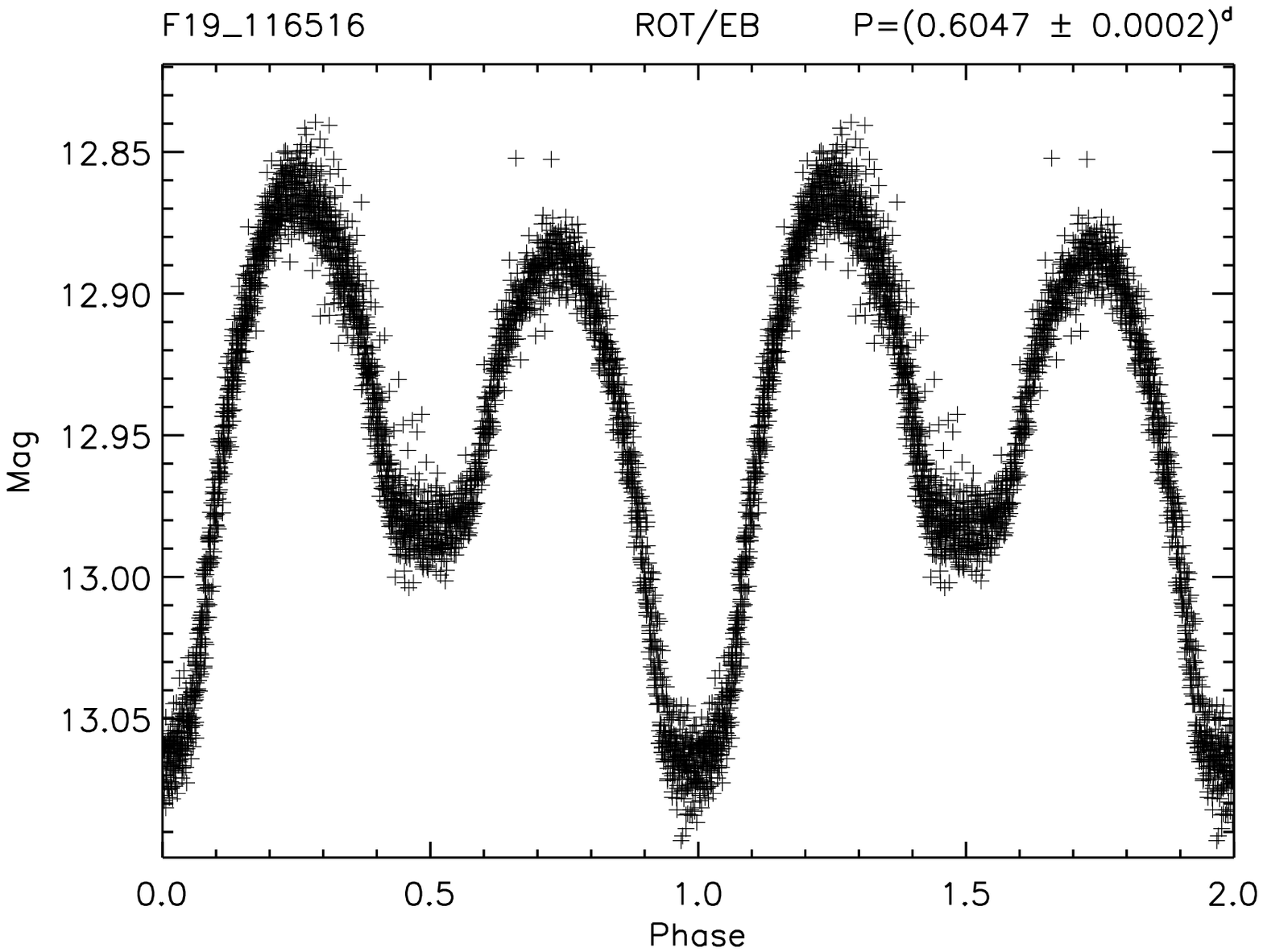} &
\includegraphics[width=\linewidth]{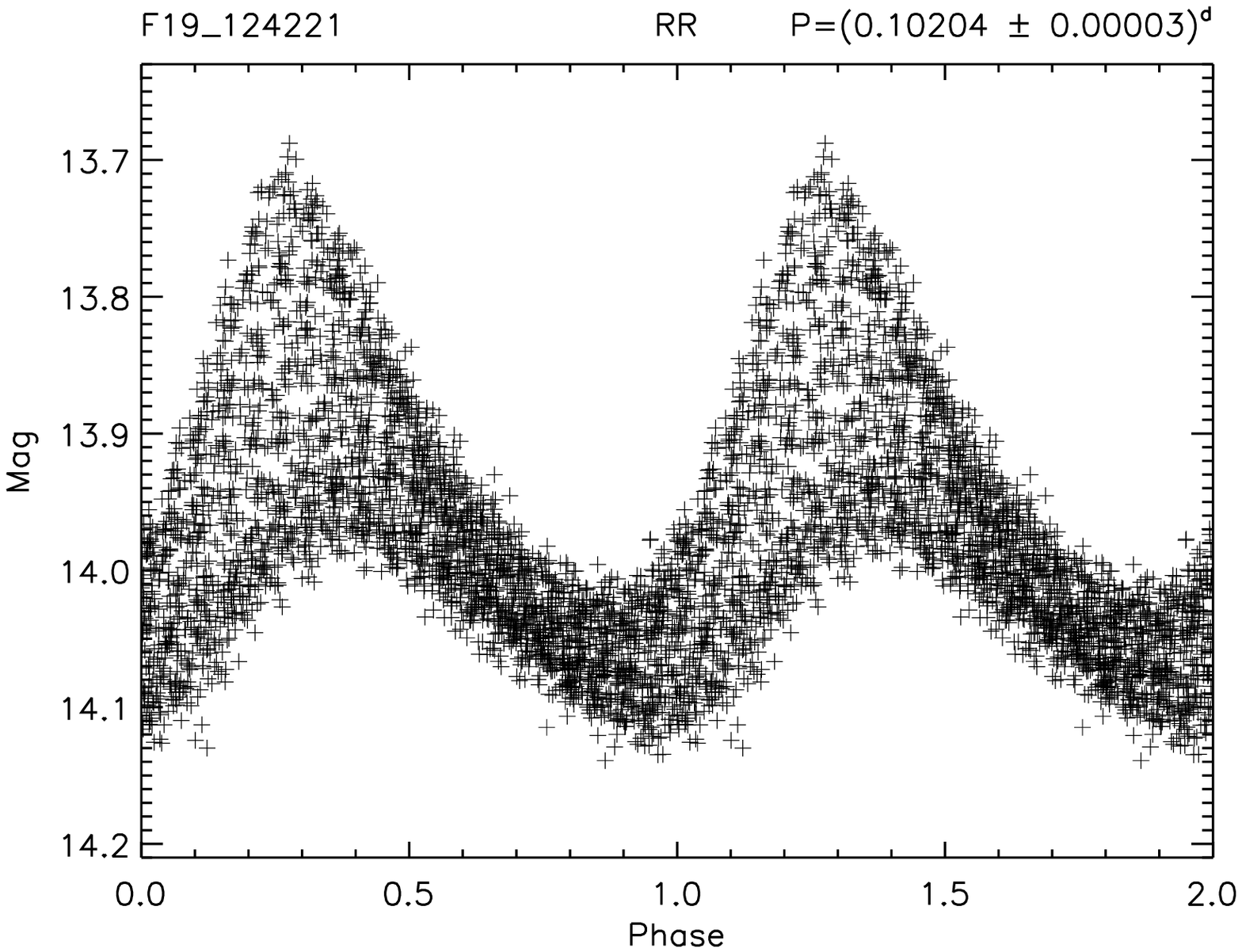} \\
\end{tabular}
\caption{Light curves of variable stars in BEST\,II target fields F17--F19 (examples). \newline (The complete figure set of 3{,}083 light curves is available in the online journal.)
\label{fig:lcs}}
\end{figure*}

A total number of 17 variable stars contained in the BEST\,II data sets F17--F19 were previously known. Table~\ref{tab:varstars:known} gives their identifiers and compares periods and classifications with the corresponding reference values. The light curves of UX~Nor (F19\_046530) and NSV\,7658 (F19\_098447) are shown as examples in Figure~\ref{fig:lcs}.

For \textit{all} stars that were classified or have periods determined by previous studies, our results are in excellent agreement. For several known variables such as EM\,Nor (F19\_104459), IX\,Nor (F19\_107786), and PW\,Nor (F19\_116322), we did not find a reference to a CCD light curve, so that this survey can be considered to provide the first high-accuracy photometric measurements of these objects. In the following text, we comment some new insights for individual cases.

\paragraph{BE\,Gru (F18\_05548), AD\,Gru (F18\_08895), and KK\,Nor (F19\_088903).} For the first time, the high precision of the new data allows to constrain the Blazhko-effect for these RR variables: Within our detection limit, we find no evidence for an amplitude modulation. This finding can be used, e.g., to study the frequency of the Blazhko-effect \citep[e.g.,][]{Sodor2012} in order to gain a better understanding of this effect. For KK\,Nor, the period was first determined within this work.
\paragraph{FV\,Nor (F19\_000499).} The long-periodic variability was first suspected by \citet{Hoffleit1931} and is now clearly confirmed. 
\paragraph{UX\,Nor (F19\_046530).} This object has one of the longest periods of RRab stars and thus is intensively studied \citep{Walraven1958,Petit1960,Diethelm1983,Diethelm1986,Diethelm1990,Kwee1984,Harris1985,Petersen1986,Petersen1987,Moskalik1993,Sandage1994,Feuchtinger1996}. The high accuracy of the new data can help to better constrain theoretical models of these pulsations \citep[see, e.g.,][]{Feuchtinger1996}.
\paragraph{NSV\,7658 (F19\_098447).} This object was only suspected to be variable by \citet{Luyten1938}. Its long-term variability is now clearly confirmed. 
\paragraph{IX\,Nor (F19\_107786).} The Mira nature of IX\,Nor is well compatible with our measurements. However, a period of its variability cannot be given since the observations do not cover a brightness maximum.
\paragraph{PW\,Nor (F19\_116322).} The Mira classification for PW\,Nor is only suspected by GCVS, but is also well supported by our measurements; they show a variation from $\sim$\,12.3 to~9.8\,mag during a half period lasting approx.~140\,days.

\subsection{Catalog of New Variables}\label{sec:catalog}
Information on all known, newly detected, and suspected variable stars identified within the three BEST\,II data sets is provided in a star catalog. Table~\ref{tab:varcat} shows a small extract for guidance regarding its form and content. The complete catalog in a machine-readable format as well as light curves for each star are available in the online journal. Photometric data and finding charts are available upon request.

\begin{deluxetable}{lcccccrccrcc}
\rotate
\tablewidth{0cm}
\setlength{\tabcolsep}{0.04in} 
\tabletypesize{\scriptsize} 
\tablecaption{Catalog of Variable Stars in Three Southern BEST\,II Target Fields (Selection)\label{tab:varcat}}
\tablehead{
\colhead{BEST\,II} & \colhead{flag} & \colhead{2MASS} & \colhead{$\alpha(J2000.0)$} & \colhead{$\delta(J2000.0)$} & 
\colhead{$R_B$ [mag]} &\colhead{$T_0$ [rHJD]} & \colhead{$P$ [d]} & \colhead{$A$ [mag]} & \colhead{$\gamma$ [\%]} & \colhead{Type} & \colhead{Other names}
}
\startdata
\cutinhead{F17}
F17\_00273  &       &14190561-5357064 & $14^h19^m05.6^s$&$-53^\circ57'06.4''$ & 14.12 &  -45.986 &          $12.0 \pm 0.3$ & $0.024 \pm 0.008$ &      1 &       ROT &\\
F17\_00277  &       &14184206-5410596 & $14^h18^m42.1^s$&$-54^\circ10'59.8''$ & 14.40 &  -52.708 &              $19 \pm 2$ &   $0.03 \pm 0.02$ &      0 &       ROT &\\
F17\_00346  &   $s$ &14194815-5332206 & $14^h19^m48.1^s$&$-53^\circ32'20.4''$ & 14.59 & $\cdots$ &                $\cdots$ &          $\cdots$ &      0 &        LP &\\
F17\_00411  &   $s$ &14184782-5408336 & $14^h18^m47.8^s$&$-54^\circ08'33.7''$ & 12.77 &  -35.920 &          $13.5 \pm 0.3$ &   $0.03 \pm 0.02$ &      0 &       ROT &\\
F17\_00448  &       &14195281-5330186 & $14^h19^m52.8^s$&$-53^\circ30'18.6''$ & 16.66 &  -57.366 &     $0.3055 \pm 0.0002$ &   $0.21 \pm 0.07$ &      3 &        EW &\\
F17\_00521  &       &14194720-5334110 & $14^h19^m47.2^s$&$-53^\circ34'10.9''$ & 14.39 & $\cdots$ &                $\cdots$ &          $\cdots$ &     14 &        SR &\\
\cutinhead{F18}
F18\_00214  &       &22471763-4420583 & $22^h47^m17.7^s$&$-44^\circ20'58.6''$ & 13.26 &   72.444 &          $12.8 \pm 0.4$ & $0.032 \pm 0.009$ &      0 &       VAR &\\
F18\_10917  &       &22543478-4418014 & $22^h54^m34.8^s$&$-44^\circ18'01.4''$ & 14.02 &   63.543 &       $0.642 \pm 0.003$ &   $0.04 \pm 0.03$ &      0 &        RR &\\
\cutinhead{F19}
F19\_000002  &   $c$ &                 & $16^h31^m57.2^s$&$-56^\circ46'40.0''$ & 14.24 &  281.119 &   $0.51944 \pm 0.00009$ &   $0.59 \pm 0.03$ &$\cdots$&        EB &\\
F19\_000271  &       &16315167-5607312 & $16^h31^m51.7^s$&$-56^\circ07'31.1''$ & 14.35 &  299.145 &              $64 \pm 7$ &   $0.08 \pm 0.02$ &      9 &       VAR &\\
F19\_000396  &       &16315404-5636039 & $16^h31^m53.9^s$&$-56^\circ36'04.7''$ & 12.79 &  293.912 &     $7.8905 \pm 0.0001$ &   $0.16 \pm 0.01$ &      8 &        EA &\\
F19\_000433  &   $c$ &16315321-5631219 & $16^h31^m53.2^s$&$-56^\circ31'22.7''$ & 12.31 &  475.611 &              $69 \pm 4$ &   $0.50 \pm 0.05$ &      5 &       CEP &\\
F19\_000441  &       &16315078-5608431 & $16^h31^m50.8^s$&$-56^\circ08'42.8''$ & 13.42 &  290.725 &          $20.2 \pm 0.8$ &   $0.03 \pm 0.02$ &     12 &       ROT &\\
F19\_000499  &  $ck$ &16315094-5614576 & $16^h31^m51.0^s$&$-56^\circ14'57.9''$ & 12.60 & $\cdots$ &                $\cdots$ &          $\cdots$ &      0 &        LP &FV Nor\\
F19\_000546  &       &16314493-5527070 & $16^h31^m44.9^s$&$-55^\circ27'07.0''$ & 15.52 &  280.778 &   $0.24814 \pm 0.00007$ &     $0.3 \pm 0.2$ &     59 &   EW/DSCT &\\
F19\_000944  &       &                 & $16^h31^m44.9^s$&$-55^\circ44'46.8''$ & 15.47 &  280.712 &     $0.5124 \pm 0.0004$ &   $0.07 \pm 0.05$ &     78 &        EA &\\
F19\_041783  &   $c$ &16280785-5524442 & $16^h28^m07.8^s$&$-55^\circ24'44.3''$ & 15.07 &  281.033 &     $0.3947 \pm 0.0002$ &   $0.08 \pm 0.05$ &     46 &        EW &F19\_041855\\
F19\_041855  &   $c$ &16280761-5524508 & $16^h28^m07.6^s$&$-55^\circ24'50.9''$ & 15.12 &  281.029 &     $0.3947 \pm 0.0003$ &   $0.07 \pm 0.04$ &     72 &        EW &F19\_041783\\
\enddata
\tablecomments{Table~\ref{tab:varcat} is published in its entirety in the electronic edition of the Astronomical Journal. A portion is shown here for guidance regarding its form and content.\\
Stars have been matched with the closest 2MASS object within a radius of 2 arcseconds. Coordinates are given for epoch J2000.0 and were derived by an astrometric match of CCD to UCAC3 coordinates. Given magnitudes are instrumental and reflect the CCD sensitivity, as observations have been obtained without filter. Overlapping apertures of neighboring stars can lead to contaminated light curves -- such cases are marked with a '$c$' flag. Suspected variables are marked by '$s$', whereas previously known variables are marked by '$k$' (see also Section~\ref{sec:knownvars} and Table~\ref{tab:varstars:known}). Amplitudes $A$ and ephemerides (variability period $P$ and times of minimum brightness $T_0$, given in rHJD=HJD-2455000) are the results of the AoV algorithm. The contamination parameter $\gamma$ approximates the fraction of light in the target's aperture coming from other objects than the target itself (see text).}
\end{deluxetable}

Each star is identified by instrumental coordinates, as well as an internal BEST\,II and 2MASS ID (if available); the latter refers to the closest 2MASS \citep{Skrutskie2006} object within $2''$ distance from the respective BEST\,II coordinates. Instrumental magnitudes $R_B$ are obtained without filter and should thus only be used for a broad approximation (cf.~Section~\ref{sec:reduction}). Ephemerides and amplitudes of variability are given based on the results of the AoV algorithm (except for cases that were adjusted as a result of visual inspection, see Section~\ref{sec:variability}). No ephemerides and amplitudes are given for long periodic (LP) classified variables.

Contaminated light curves are marked with a~``$c$'' flag in the catalog. If the origin of variation can clearly be assigned to one of the overlapping stars due to a sufficient angular separation and/or brightness difference, only one object is presented in the catalog. However, in 43 cases, two objects are too close to each other and are thus both presented as variables (marked with the corresponding ID under "other names" in Table~\ref{tab:varcat}, but not counted twice in Tables~\ref{tab:varstars} and~\ref{tab:varstartypes}). Observations at higher angular resolution are needed to constrain the true origin of variability.

\subsection{Catalog Characterization}\label{sec:results:sens}
The new variables of this study will increase the number of variables listed in the VSX catalog (246{,}007 as of 14/08/2013, including suspected) by~1.2\%. For comparison, Table~\ref{tab:surveycomp} shows the volume, yield, and magnitude range of this work, previous BEST/BEST\,II studies, and other large photometric surveys.

\begin{deluxetable}{l@{}rrccp{6cm}}
\tablecaption{Variable star detection yield in comparison to other surveys.\label{tab:surveycomp}}
\tabletypesize{\footnotesize} 
\tablehead{\colhead{Project} & \colhead{$N_\star$} & \colhead{$N_\textrm{var}$} & \colhead{$N_\textrm{var}/N_\star$} & \colhead{Mag} & \colhead{References}}
\startdata
\multicolumn{6}{l}{\textsc{BEST/BEST\,II}}\\\hline
BEST                & 121{,}811 &     335 & 0.28\% & 11--15 & \citet{Karoff2007,Kabath2007,Kabath2008,Pasternacki2011} \\
BEST\,II (CoRoT)    & 218{,}809 & 1{,}107 & 0.51\% & 11--17 & \citet{Kabath2009,Kabath2009a,Fruth2012} \\
BEST\,II (F17--F19) & 209{,}070 & 3{,}040 & 1.45\% & 11--17 & This work \\\hline
\multicolumn{6}{l}{\textsc{Other Ground-based Surveys}}\\\hline
UNSW     &        87{,}000 &       850 & 0.98\% &  8--14 & \citet{Christiansen2008} \\
HAT 199  &        98{,}000 &   1{,}617 & 1.65\% &  8--14 & \citet{Hartman2004} \\
EROS II  &   1{,}913{,}576 &   1{,}362 & 0.07\% & 11--17 & \citet{Derue2002} \\
ASAS-2   &       140{,}000 &   3{,}800 & 2.71\% &  8--13 & \citet{Pojmanski2000} \\
ASAS-3   & $1.7\cdot 10^7$ &  50{,}099 & 0.29\% &  8--14 & \citet{Paczy'nski2006} \\
OGLE-II  & $1.65\cdot 10^7$&  68{,}194 & 0.41\% & $\leq$\,20 & \citet{Zebrun2001} \\
OGLE-III &   $2\cdot 10^8$ & 193{,}000 & 0.10\% & 12--20 & \citet{Soszynski2008,Soszynski2011} \\
\hline
\multicolumn{6}{l}{\textsc{Space-based Surveys}}\\\hline
CoRoT (first 4 fields) &  39{,}659 && $\sim$\,40\% & 12--16 & \citet{Debosscher2009} \\
Kepler                 & 156{,}000 &&              & 9--16  & \citet{Borucki2011} \\
\multicolumn{2}{l@{}}{\dotfill\ \textit{(quasi-)periodic}}  &         & 16\%  && \citet{McQuillan2012} \\
\multicolumn{2}{l@{}}{\dotfill\ \textit{eclipsing binaries}}& 1{,}879 & 1.2\% && \citet{Slawson2011} \\
\enddata
\vspace{-5mm}
\tablecomments{The table gives the number of surveyed stars $N_\star$, the number of found variables $N_\textrm{var}$, and the corresponding ratio $N_\textrm{var}/N_\star$ for BEST/BEST\,II publications, this work, and selected references of important variable star surveys. If noted in the publication, $N_\textrm{var}$ here includes known and new detections, i.e., the whole detection yield of a given survey. The column "Mag" gives each survey's approximate magnitude range.}
\end{deluxetable}

The detection yield of photometric surveys is subject to various systematic differences. Most importantly, these include the photometric precision, monitored magnitude range and FOV, the time span and duty cycle of observations, and the applied analysis techniques and selection criteria; for a discussion, see also \citealt{Tonry2005}. For ground-based projects, the detection yield typically ranges between 0.1\% (OGLE-III) and 2.7\% (ASAS-2), which compares well to the overall fraction of 1.2\% variables (1.5\% including suspected) identified in this study. However, the increased precision of space-based surveys indicates a much larger fraction of stars to be variable -- \citet{McQuillan2012} detected clear periodic or quasi-periodic behaviour for 16\% of stars in \textit{Kepler} \citep{Borucki2010} data, and \citet{Debosscher2009} estimated at least 40\% of CoRoT light curves to be variable. 

In the following text, the sensitivity of this study is investigated as a function of its three most important limitations, namely, the magnitude range, the SNR, and the period range.

\subsubsection{Magnitude Range and Variable Fraction}
In order to evaluate the completeness of our study as a function of apparent brightness, we compared the number of observed stars against the GSC2.2 catalog \citep{Lasker2008}, which resembles our results the best within the given magnitude range among several catalogs tested. Differences between the instrumental photometric system $R_B$ and the $R$ band of GSC2.2 were corrected for by subtracting the average difference~$\left<\Delta m\right>$ to obtain 
\begin{equation}\label{eq:RBshifted}
R'_B=R_B-\left<\Delta m\right>, 
\end{equation}
whereby $\Delta m=R_B-R$. 

\begin{figure}[htc]\centering
  \includegraphics[width=\linewidth]{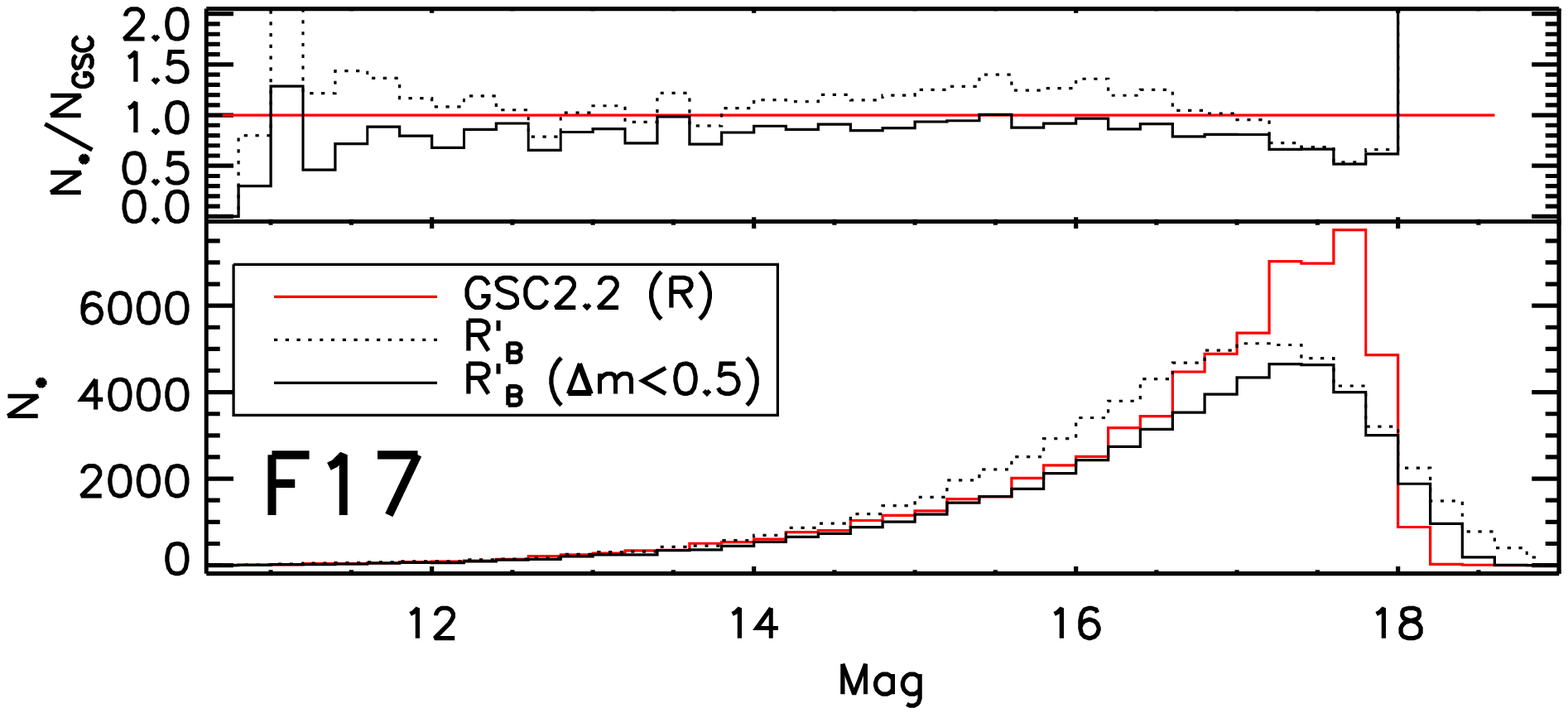}
  \includegraphics[width=\linewidth]{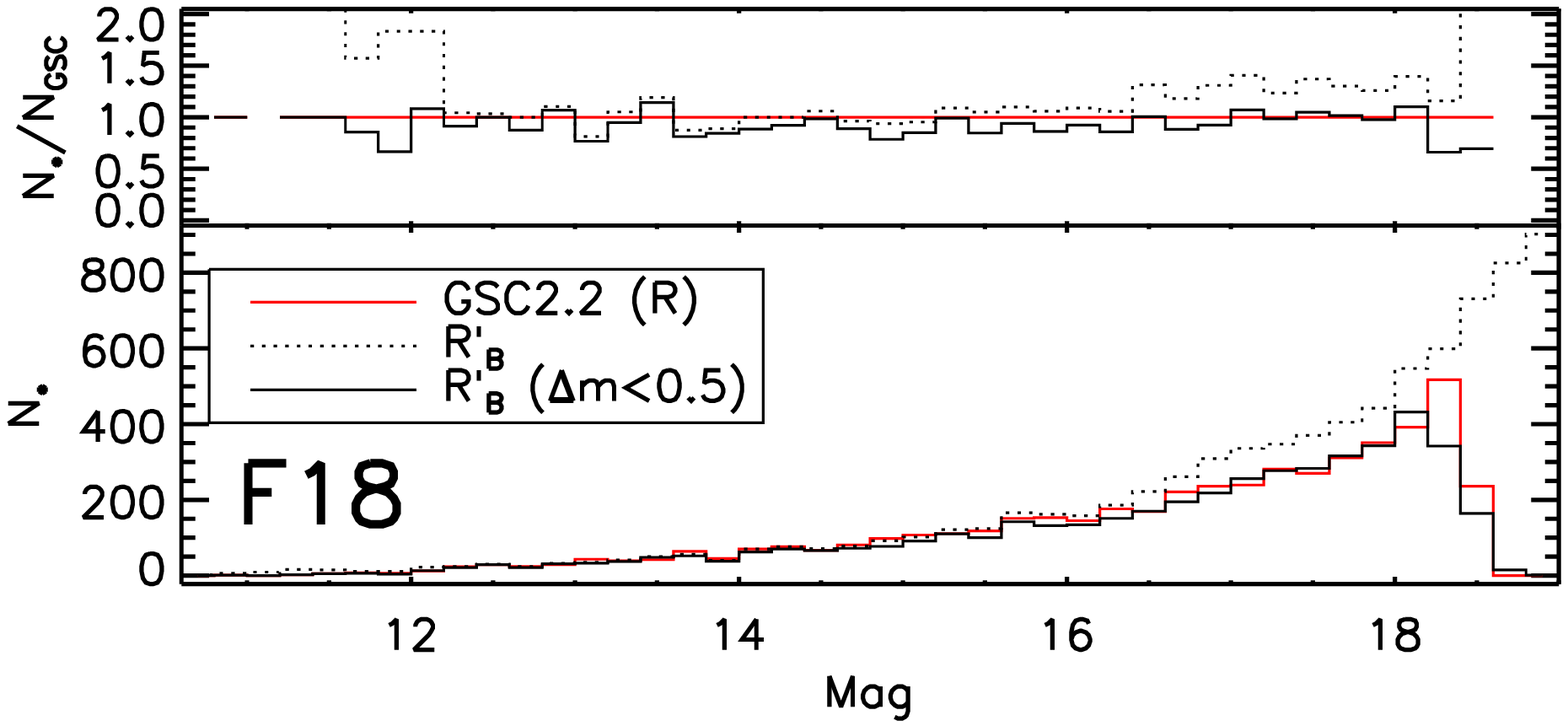}
  \includegraphics[width=\linewidth]{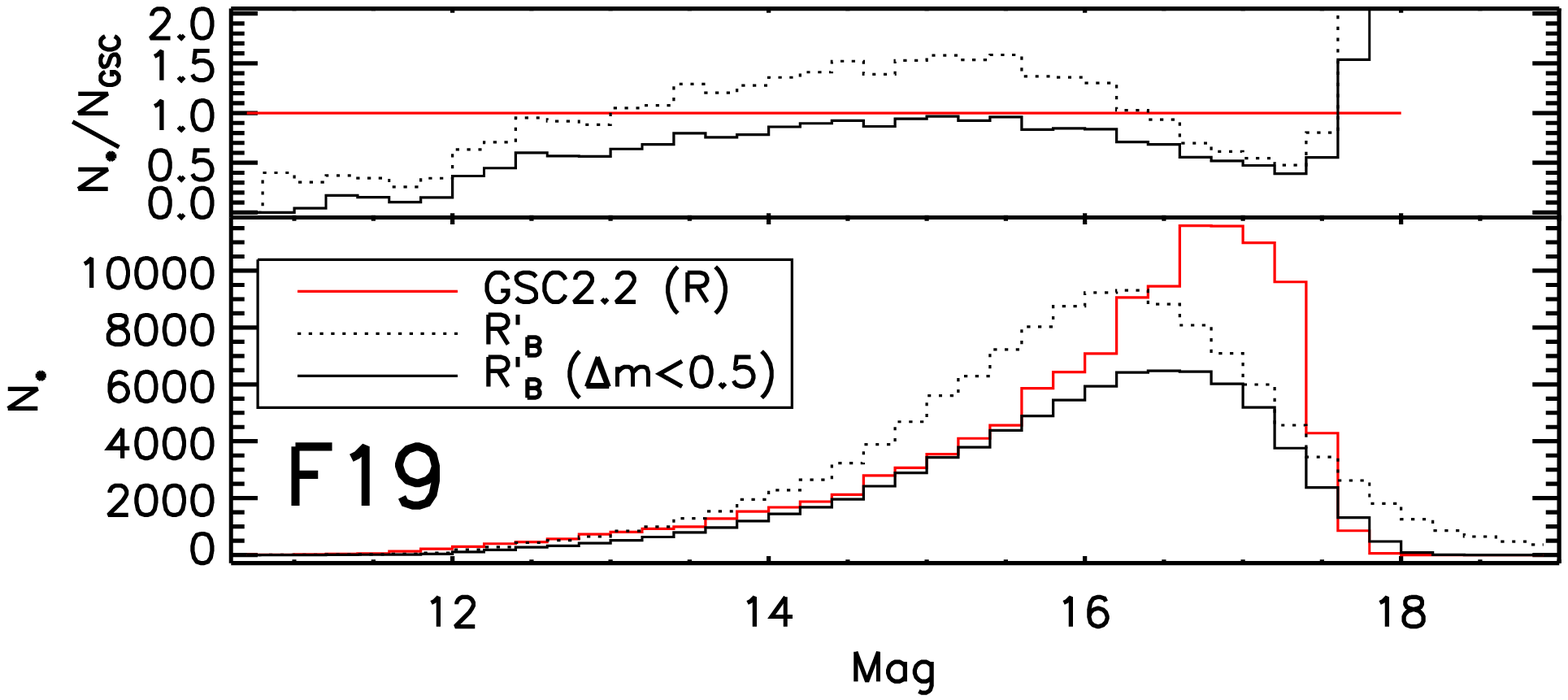}    
\figcaption{Magnitude histogram for each BEST\,II target field (in bins of 0.2\,mag). The lower panel shows the number of stars $N_\star$ as a function of the GSC2.2 catalog $R$~magnitude  ($N_\star\equiv N_{\textrm{GSC}}$; red), BEST\,II instrumental magnitude $R'_B$ (Eq.~\ref{eq:RBshifted}; black dotted), and of BEST\,II stars with \mbox{$\left|\Delta m\right|<0.5$\,mag} (black). The upper panel shows the corresponding ratio $N_\star/N_{\textrm{GSC}}$.
\label{fig:hist:mag1}}
\end{figure}
Figure~\ref{fig:hist:mag1} compares the number of catalog stars (within bins of 0.2\,mag) with the respective numbers for each target field of this work. 
The latter is shown both for all stars and a reduced sample having $\left|\Delta m\right|<0.5$\,mag, since large deviations may indicate a systematic disagreement, e.g., due to different angular resolutions. The comparison shows a very good agreement between catalog and survey stars ($\gtrsim$\,80\% completeness) within the magnitude range of \mbox{$11\lesssim R\lesssim 17$} for the two data sets F17 and F18. For target field F19, a similarly good agreement is confined to the range of \mbox{$13\lesssim R\lesssim 16$}, since most stars with $R\lesssim 13$ are saturated (due to the longer exposure time compared to F17 and F18), and stars with $R\gtrsim 16$ are more strongly affected by crowding in this dense target field.

\begin{figure*}[p]\centering
  \includegraphics[width=.49\linewidth]{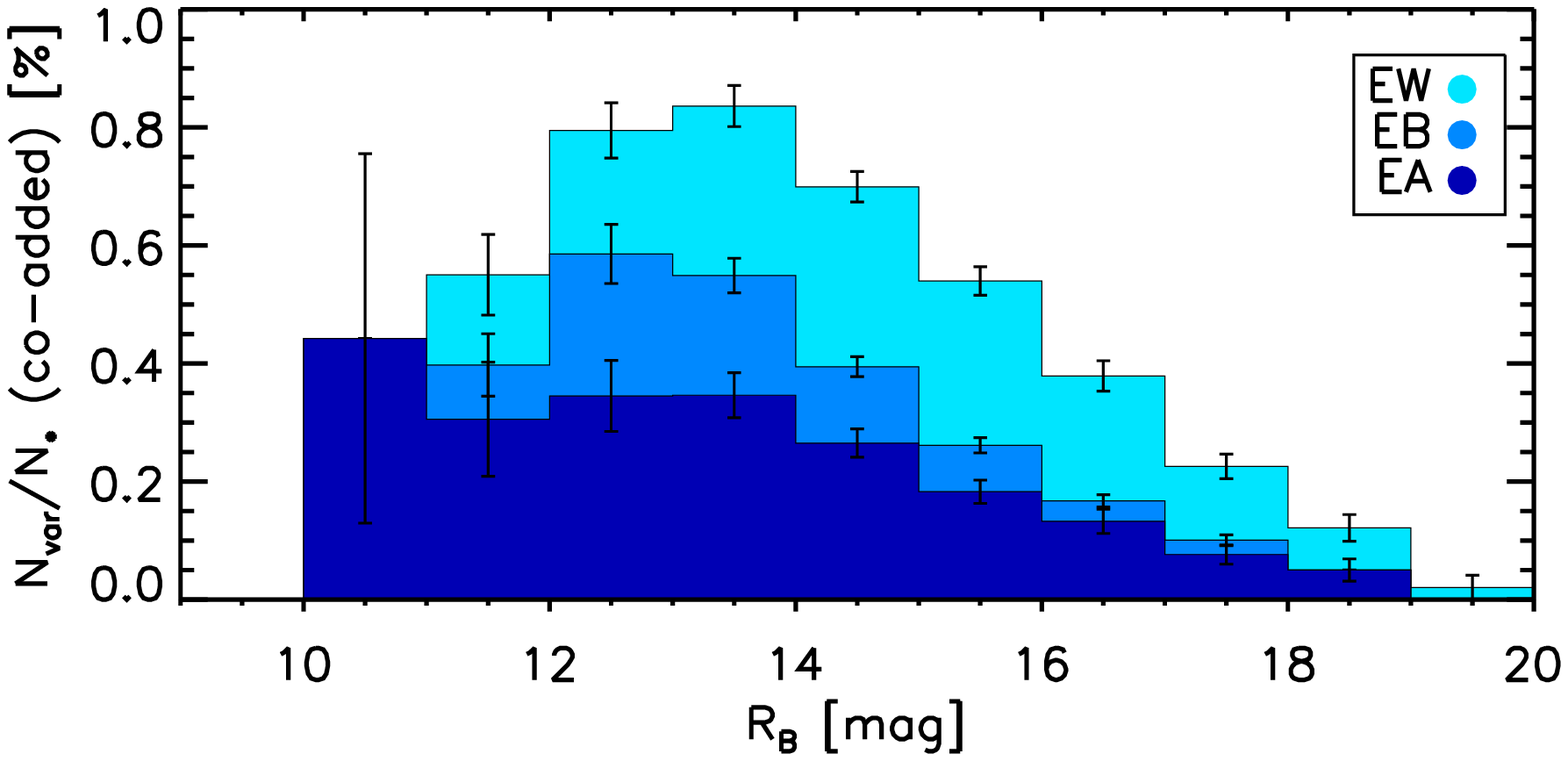}
  \includegraphics[width=.49\linewidth]{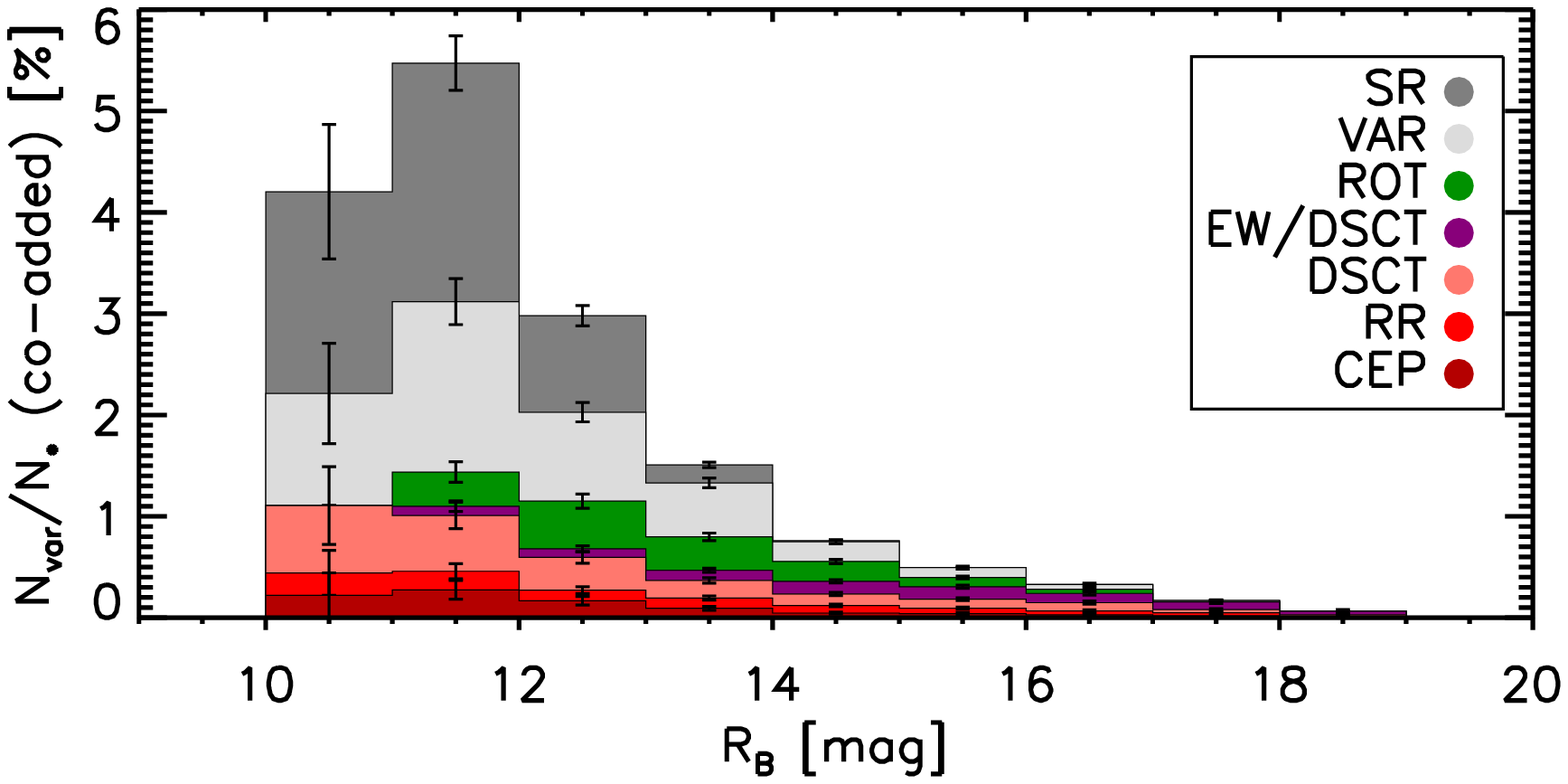}
\figcaption{Ratio of the variable star count $N_\textrm{var}$ (including suspected variables) and the total number $N_\star$ of field stars as a function of $R_B$ within bins of 1\,mag. The left panel shows eclipsing binaries, while the right panel shows all remaining major classes. Bins are co-added, i.e., one complete bar refers to the total fraction of \textit{all} classes. Vertical bars indicate the uncertainties of $N_\textrm{var}/N_\star$ (assuming a Poisson distribution for both $N_\textrm{var}$ and $N_\star$).
\label{fig:hist:mag2}}
\end{figure*}

Figure~\ref{fig:hist:mag2} shows the number of variables and suspected variables $N_\textrm{var}$ identified within this work as a fractional ratio of all stars $N_\star$ per magnitude bin separated into binaries and other types. Naturally, the detection yield strongly decreases towards fainter magnitudes as the light curve precision decreases (cf.~Figure~\ref{fig:rmsplot}). It peaks at $0.836(4)\%$ for the binaries ($R_B\in\left[13,14\right]$) and at $5.47(8)\%$ for other types of variability ($R_B\in\left[11,12\right]$). These values can be considered an overall lower limit of the real fraction of variable stars in our target fields, assuming that the physical dependency on the apparent brightness is small. Within the most sensitive magnitude range of this survey at \mbox{$11\leq R_B\leq 17$}, we identified an average fraction of $0.67(3)\%$ eclipsing binaries. This compares well to the results of other photometric surveys such as OGLE, which identified an eclipsing binary fraction of \mbox{$\sim$\,0.5--0.6\%} ($I\in\left[14,17\right]$; \citealt{Graczyk2011}, cf.~their Figure~2). However, the fraction is significantly smaller compared to space surveys: \citet{Slawson2011} reported $\sim$\,2.0\% eclipsing binaries at comparable galactic latitudes in the \textit{Kepler} field. If the latter approximately resembled the physical content in our target fields, the fraction of eclipsing binaries identified in this survey would correspond to a completeness of about one third for $R_B\in\left[11,17\right]$.

\begin{figure*}[p]\centering
  \includegraphics[width=.49\linewidth]{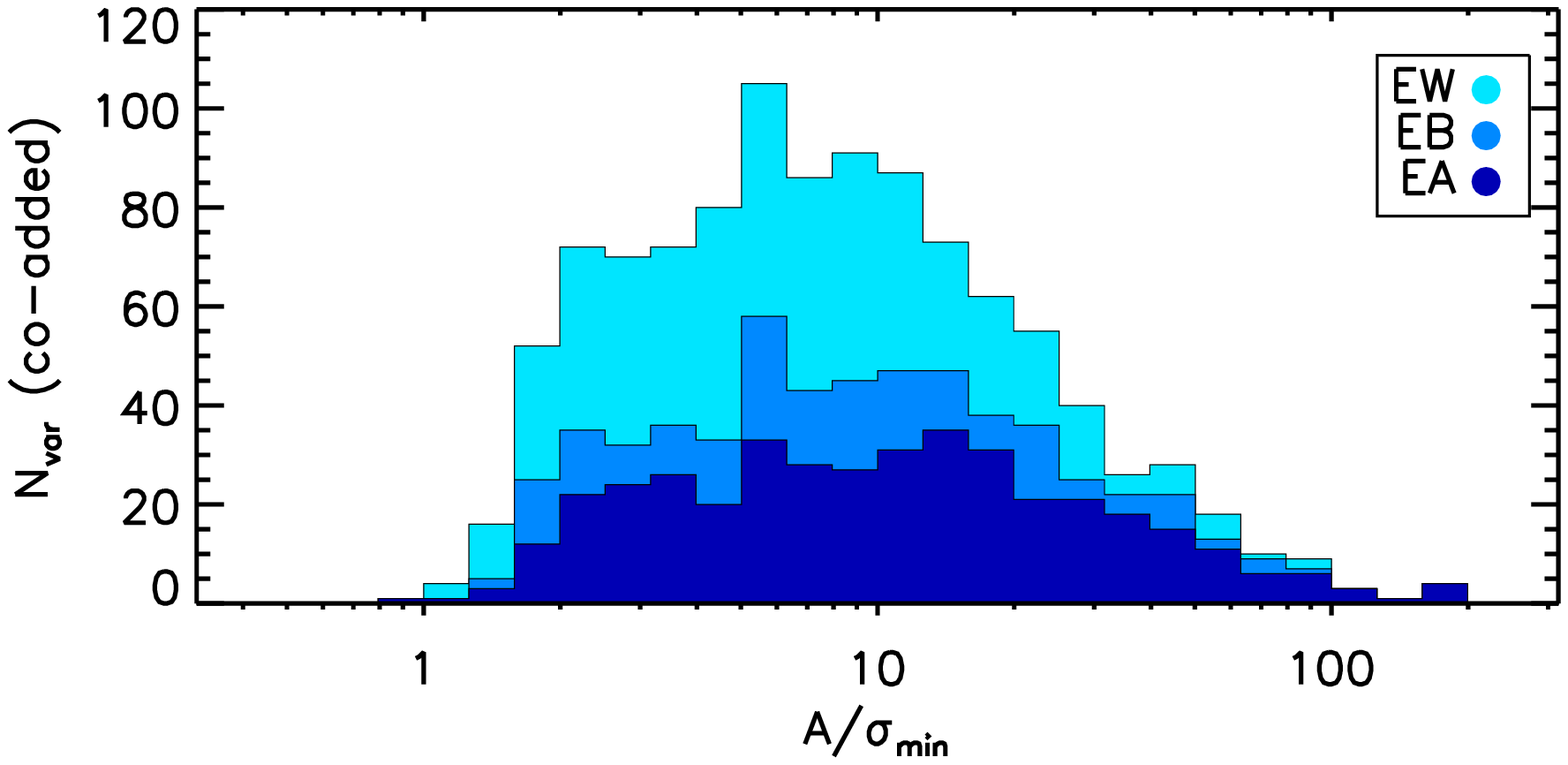}
  \includegraphics[width=.49\linewidth]{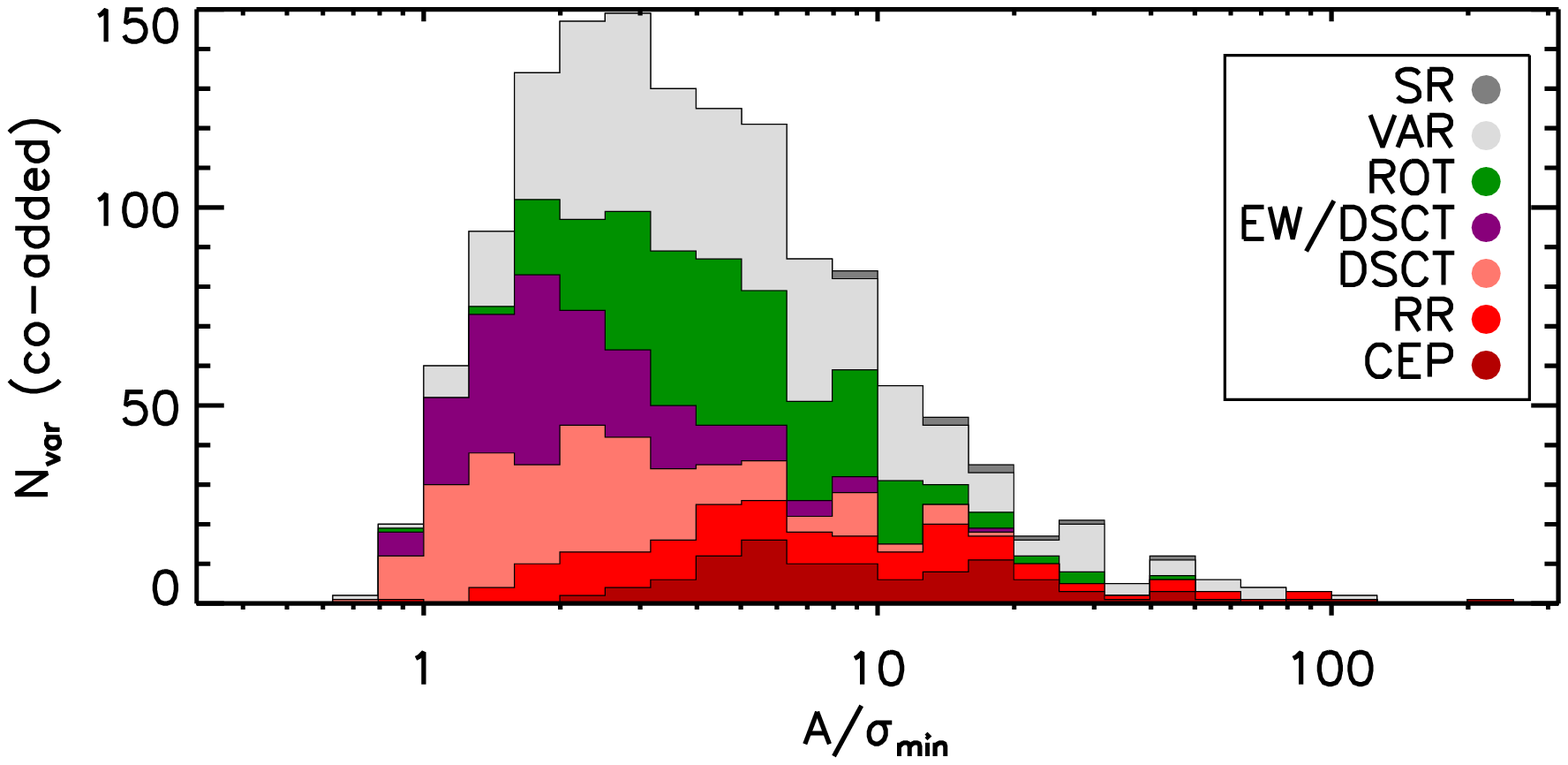}
\figcaption{Variable star count $N_\textrm{var}$ (including suspected) as a function of $\textrm{SNR}=A/\sigma_\textrm{min}$ in unbinned data (bin size $0.1\log_{10}(A/\sigma_\textrm{min})$). The amplitudes $A$ are taken from Table~\ref{tab:varcat}, while $\sigma_\textrm{min}$ refers to the minimum noise level encountered at each respective magnitude (red line in Figure~\ref{fig:rmsplot}). Types, colors and co-adding as in Figure~\ref{fig:hist:mag2}. \label{fig:hist:SNR}}
\end{figure*}

\begin{figure*}[p]\centering
  \includegraphics[width=.49\linewidth]{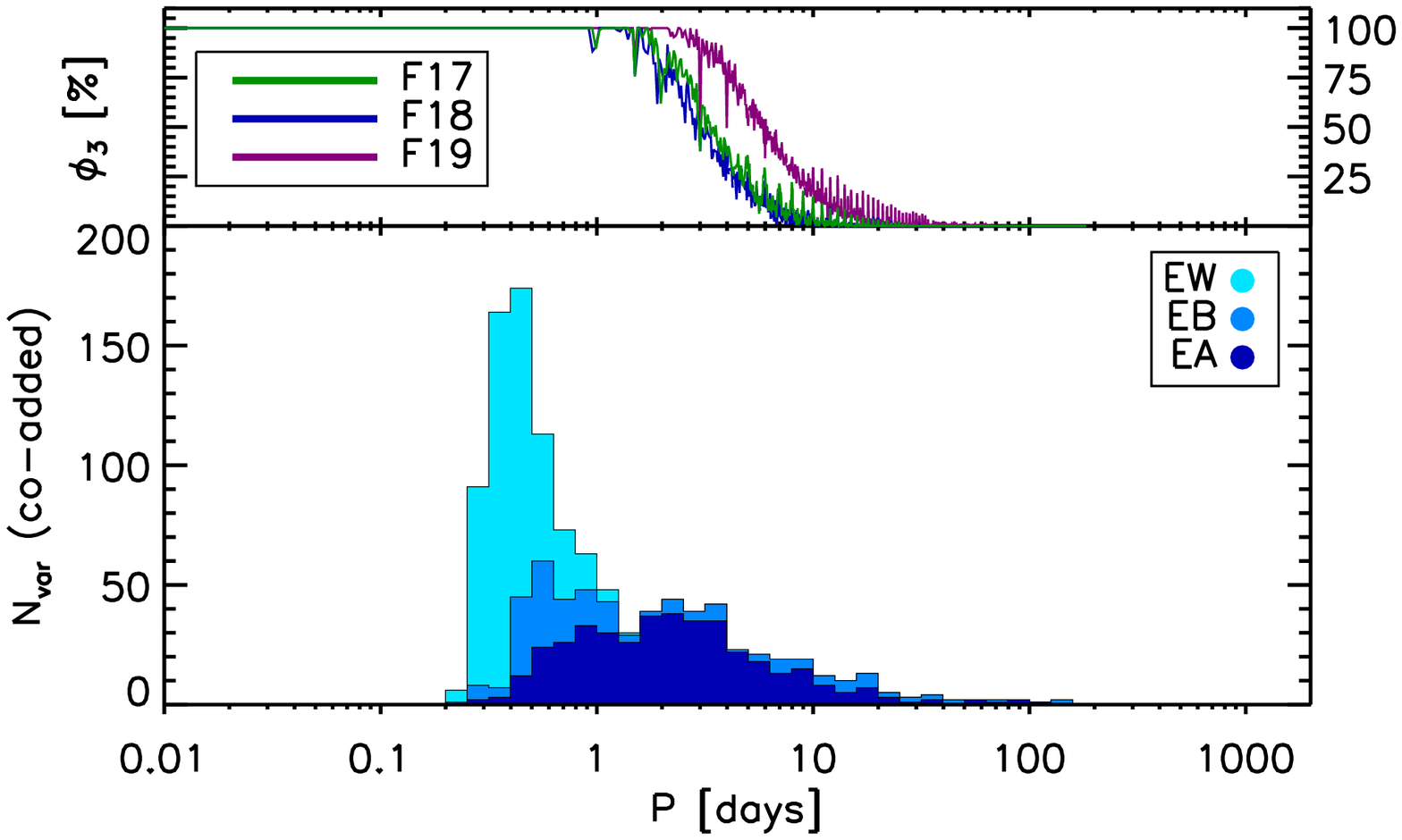}
  \includegraphics[width=.49\linewidth]{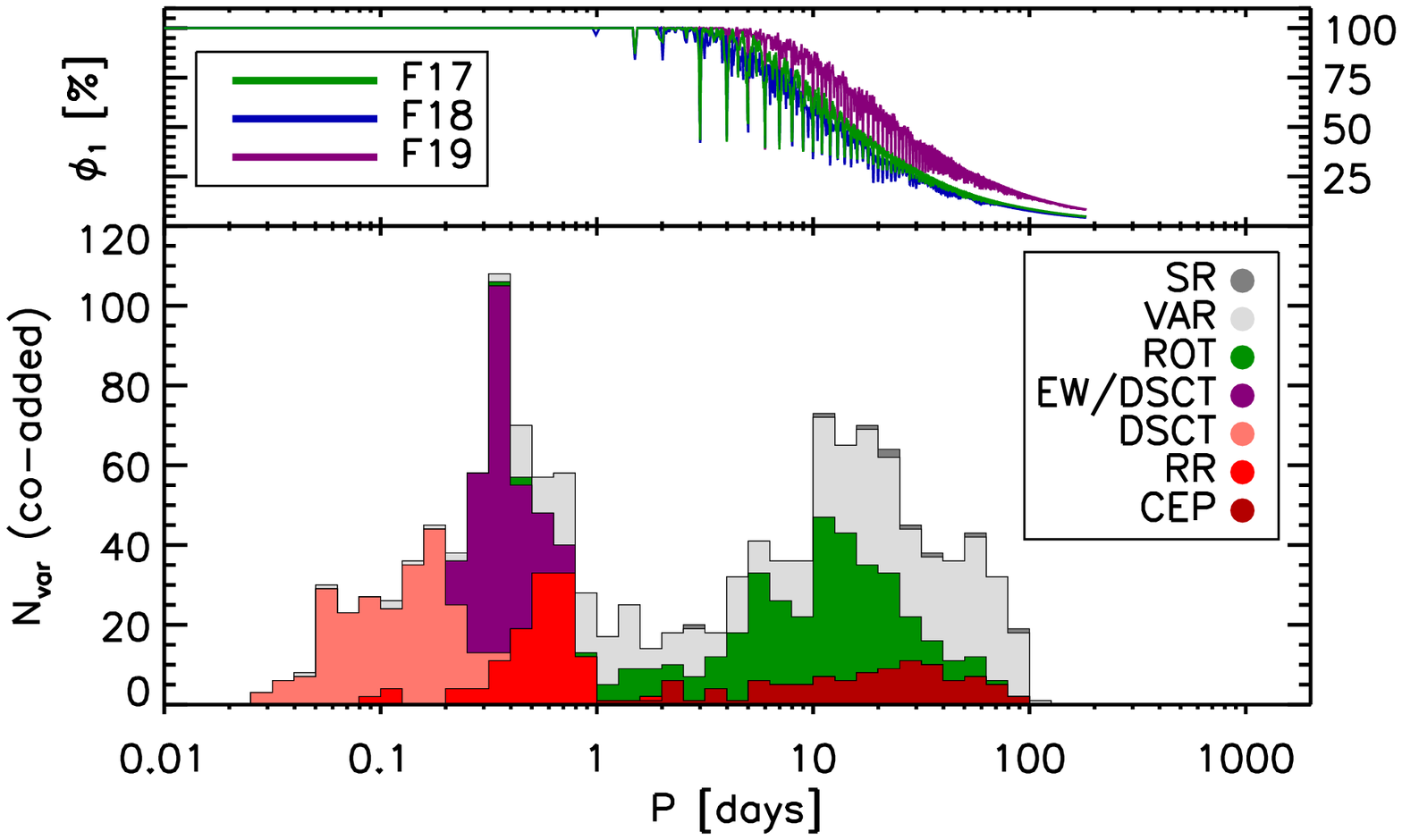}
\figcaption{Variable star count $N_\textrm{var}$ (including suspected) as a function of the period of variability~$P$ (see Table~\ref{tab:varcat}; bin size $0.1\log_{10}(P/\textrm{days})$). The upper section of both panels shows the single and triple phase coverage, $\phi_1$ and $\phi_3$, respectively, calculated from observing times of each field (cf.~Figure~\ref{fig:best2obs}). Colors and co-adding as in Figure~\ref{fig:hist:mag2}.\label{fig:hist:percov}}
\end{figure*}

\subsubsection{SNR Limit}
In order to evaluate our detection threshold, the amplitude $A$ of each variable star is compared with the minimum noise level $\sigma_\textrm{min}(R_B)$, which is determined for each field from a fit to the lower noise limit in the ($R_B$, $\sigma$)-diagram (cf.~Figure~\ref{fig:rmsplot}; after an analytical expression by \citealt{Newberry1991}, Eq.~12).

Figure~\ref{fig:hist:SNR} shows a histogram of the SNR $A/\sigma_\textrm{min}$ for eclipsing binaries and other types of variables. For the eclipsing binaries, the detection yield strongly increases for $A/\sigma_\textrm{min}\gtrsim 2$, while some types with a more sinusoidal variation such as DSCT can already be detected at $A/\sigma_\textrm{min}\gtrsim 1$. Note that $\sigma_\textrm{min}$ here refers to \textit{unbinned} data, although a large number of single measurements effectively contribute to each detection, thus increasing the SNR significantly.

\subsubsection{Period Range}
The duty cycle of any photometric survey significantly constrains its detection efficiency, in particular for variability at large time scales. Figure~\ref{fig:hist:percov} shows the number of variables detected in this work as a function of their period. For comparison, the observational phase coverage~$\phi_1(P)$ is shown for each target field. Since the detection of detached eclipsing binaries (EA) usually requires an observation of three or more of the relatively short eclipses, the right panel of Figure~\ref{fig:hist:percov} shows the phase coverage~$\phi_3(P)$ of at least three (infinitesimally short) events.

Figure~\ref{fig:hist:percov} shows that our detection efficiency is largely confined to periods of $P\lesssim 10$\,days in the case of eclipsing binaries, while the more regular periodic variability of other types can be detected up to $P\lesssim 100$\,days. Qualitatively, the decrease in efficiency well follows the single and triple observational phase coverage, respectively. Furthermore, the period ranges of several distributions reflect their class definition (e.g., DSCT variables with $P\leq0.2$--0.3\,days). The observed peak of contact binaries (EW) at $P\approx0.3$--0.4\,days agrees well with recent findings of other surveys (e.g., ASAS, \citealt{Paczy'nski2006}; \textit{Kepler}, \citealt{Slawson2011}; OGLE, \citealt{Pietrukowicz2013}).

\section{Summary}\label{sec:summary}
A BEST\,II photometric survey of 209{,}070 stars within three Southern target fields revealed 2{,}423 variable objects (from which 17 were previously known) and an additional 617 stars which are suspected to be variable. The underlying stellar sample is most complete ($\gtrsim$\,80\%) within the magnitude range of \mbox{$R\in\left[13,16\right]$} and the survey is most sensitive to detect eclipsing binaries up to periods of $\sim$\,10\,days and other types up to $\sim$\,100\,days. The average fraction of $0.67(3)\%$ eclipsing binaries at magnitudes \mbox{$R\in\left[11,17\right]$} agrees well with other ground-based surveys and indicates a detection completeness of about one third in comparison to the more sensitive space surveys.

All information that is available from our photometric measurements of variable stars is presented in a large catalog, including the type and/or periodicity of the variability (if determinable). The catalog encompasses a number of objects that are very interesting for further astrophysical studies beyond the scope of this paper.

\acknowledgments
\textbf{Acknowledgments.} This work was funded by Deutsches Zentrum f\"ur Luft- und Raumfahrt and partly by the Nordrhein-Westf\"alische Akademie der Wissenschaften. Our research made use of the 2MASS, USNO-A2, NOMAD, GSC2.2, and GCVS catalog, the AAVSO variable star search index and the SIMBAD database, operated at CDS, Strasbourg, France.

\bibliographystyle{apj}
\bibliography{ms}

\end{document}